\documentclass[twocolumn]{aastex631}

\shortauthors{Lewin et al.}
\graphicspath{{./}{/}}

\begin{document}

\title{AGN STORM 2. VII. A Frequency-resolved Map of the Accretion Disk in Mrk~817: Simultaneous X-ray Reverberation and UVOIR Disk Reprocessing Time Lags}

\correspondingauthor{Collin Lewin}
\email{clewin@mit.edu}

\author[0000-0002-8671-1190]{Collin Lewin}
\affiliation{MIT Kavli Institute for Astrophysics and Space Research, MIT, 77 Massachusetts Avenue, Cambridge, MA 02139, USA}

\author[0000-0003-0172-0854]{Erin Kara}
\affiliation{MIT Kavli Institute for Astrophysics and Space Research, MIT, 77 Massachusetts Avenue, Cambridge, MA 02139, USA}

\author[0000-0002-3026-0562]{Aaron J.\ Barth}
\affiliation{Department of Physics and Astronomy, 4129 Frederick Reines Hall, University of California, Irvine, CA, 92697-4575, USA}

\author[0000-0002-8294-9281]{Edward M.\ Cackett}
\affiliation{Department of Physics and Astronomy, Wayne State University, 666 W.\ Hancock St, Detroit, MI, 48201, USA}

\author[0000-0003-3242-7052]{Gisella De~Rosa}
\affiliation{Space Telescope Science Institute, 3700 San Martin Drive, Baltimore, MD 21218, USA}

\author[0000-0002-0957-7151]{Yasaman Homayouni}
\affiliation{Space Telescope Science Institute, 3700 San Martin Drive, Baltimore, MD 21218, USA}
\affiliation{Department of Astronomy and Astrophysics, The Pennsylvania State University, 525 Davey Laboratory, University Park, PA 16802}
\affiliation{Institute for Gravitation and the Cosmos, The Pennsylvania State University, University Park, PA 16802}

\author[0000-0003-1728-0304]{Keith Horne}
\affiliation{SUPA School of Physics and Astronomy, North Haugh, St.~Andrews, KY16~9SS, Scotland, UK}

\author[0000-0002-2180-8266]{Gerard A.\ Kriss}
\affiliation{Space Telescope Science Institute, 3700 San Martin Drive, Baltimore, MD 21218, USA}

\author{Hermine Landt}
\affiliation{Centre for Extragalactic Astronomy, Department of Physics, Durham University, South Road, Durham DH1 3LE, UK}

\author[0000-0001-9092-8619]{Jonathan Gelbord}
\affiliation{Spectral Sciences Inc., 30 Fourth Ave. Suite 2, Burlington MA 01803}

\author[0000-0001-5639-5484]{John Montano}
\affiliation{Department of Physics and Astronomy, 4129 Frederick Reines Hall, University of California, Irvine, CA, 92697-4575, USA}

\author[0000-0003-2991-4618]{Nahum Arav}
\affiliation{Department of Physics, Virginia Tech, Blacksburg, VA 24061, USA}

\author[0000-0002-2816-5398]{Misty C.\ Bentz}
\affiliation{Department of Physics and Astronomy, Georgia State University, 25 Park Place, Suite 605, Atlanta, GA 30303, USA}

\author[0000-0001-6301-570X]{Benjamin D.\ Boizelle}
\affiliation{Department of Physics and Astronomy, N284 ESC, Brigham Young University, Provo, UT, 84602, USA}

\author[0000-0001-9931-8681]{Elena Dalla Bont\`{a}}
\affiliation{Dipartimento di Fisica e Astronomia ``G.\  Galilei,'' Universit\`{a} di Padova, Vicolo dell'Osservatorio 3, I-35122 Padova, Italy}
\affiliation{INAF-Osservatorio Astronomico di Padova, Vicolo dell'Osservatorio 5 I-35122, Padova, Italy}

\author[0000-0002-1207-0909]{Michael S.\ Brotherton}
\affiliation{Department of Physics and Astronomy, University of Wyoming, Laramie, WY 82071, USA}

\author[0000-0002-0964-7500]{Maryam Dehghanian}
\affiliation{Department of Physics, Virginia Tech, Blacksburg, VA 24061, USA}

\author[0000-0003-4503-6333]{Gary J.\ Ferland}
\affiliation{Department of Physics and Astronomy, The University of Kentucky, Lexington, KY 40506, USA}

\author[0000-0002-2306-9372]{Carina Fian}
\affiliation{Departamento de Astronom\'{i}a y Astrof\'{i}sica, Universidad de Valencia, E-46100 Burjassot, Valencia, Spain}
\affiliation{Observatorio Astron\'{o}mico, Universidad de Valencia, E-46980 Paterna, Valencia, Spain}

\author[0000-0002-2908-7360]{Michael R.\ Goad}
\affiliation{School of Physics and Astronomy, University of Leicester, University Road, Leicester, LE1 7RH, UK}

\author[0000-0002-6733-5556]{Juan V.\ Hern\'{a}ndez Santisteban}
\affiliation{SUPA School of Physics and Astronomy, North Haugh, St.~Andrews, KY16~9SS, Scotland, UK}

\author[0000-0002-1134-4015]{Dragana Ili\'{c}}
\affiliation{University of Belgrade - Faculty of Mathematics, Department of astronomy, Studentski trg 16, 11000 Belgrade, Serbia}
\affiliation{Hamburger Sternwarte, Universit\"at Hamburg, Gojenbergsweg 112, 21029 Hamburg, Germany}

\author[0000-0001-5540-2822]{Jelle Kaastra}
\affiliation{SRON Netherlands Institute for Space Research, Niels Bohrweg 4, 2333 CA Leiden, The Netherlands}
\affiliation{Leiden Observatory, Leiden University, PO Box 9513, 2300 RA Leiden, The Netherlands}

\author[0000-0002-9925-534X]{Shai Kaspi}
\affiliation{School of Physics and Astronomy and Wise observatory, Tel Aviv University, Tel Aviv 6997801, Israel}

\author[0000-0003-0944-1008]{Kirk T.\ Korista}
\affiliation{Department of Physics, Western Michigan University, 1120 Everett Tower, Kalamazoo, MI 49008-5252, USA}

\author[0000-0003-4511-8427]{Peter Kosec}
\affiliation{MIT Kavli Institute for Astrophysics and Space Research, MIT, 77 Massachusetts Avenue, Cambridge, MA 02139, USA}

\author[0000-0001-5139-1978]{Andjelka Kova\v{c}evi\'{c}}
\affiliation{University of Belgrade - Faculty of Mathematics, Department of astronomy, Studentski trg 16, 11000 Belgrade, Serbia}

\author[0000-0002-4992-4664]{Missagh Mehdipour}
\affiliation{Space Telescope Science Institute, 3700 San Martin Drive, Baltimore, MD 21218, USA}

\author[0000-0001-8475-8027]{Jake A. Miller}
\affiliation{Department of Physics and Astronomy, Wayne State University, 666 W.\ Hancock St, Detroit, MI, 48201, USA}

\author[0000-0002-6766-0260]{Hagai Netzer}
\affiliation{School of Physics and Astronomy and Wise observatory, Tel Aviv University, Tel Aviv 6997801, Israel}

\author[0000-0001-7351-2531]{Jack M.\ M.\ Neustadt}
\affiliation{Department of Astronomy, The Ohio State University, 140 W. 18th Ave., Columbus, OH 43210, USA}

\author{Christos Panagiotou}
\affiliation{MIT Kavli Institute for Astrophysics and Space Research, MIT, 77 Massachusetts Avenue, Cambridge, MA 02139, USA}

\author[0000-0003-1183-1574]{Ethan R. Partington}
\affiliation{Department of Physics and Astronomy, Wayne State University, 666 W.\ Hancock St, Detroit, MI, 48201, USA}

\author[0000-0003-2398-7664]{Luka \v{C}.\ Popovi\'{c}}
\affiliation{Astronomical Observatory Belgrade, Volgina 7, 11000 Belgrade, Serbia}
\affiliation{University of Belgrade - Faculty of Mathematics, Department of astronomy, Studentski trg 16, 11000 Belgrade, Serbia}

\author[0000-0002-9238-9521]{David Sanmartim}
\affiliation{Carnegie Observatories, Las Campanas Observatory, Casilla 601, La Serena, Chile} 

\author[0000-0001-9191-9837]{Marianne Vestergaard}
\affiliation{Steward Observatory, University of Arizona, 933 North Cherry Avenue, Tucson, AZ 85721, USA} 
\affiliation{DARK, The Niels Bohr Institute, University of Copenhagen, Jagtvej 155, DK-2200 Copenhagen, Denmark}

\author[0000-0003-1810-0889]{Martin J.\ Ward}
\affiliation{Centre for Extragalactic Astronomy, Department of Physics, Durham University, South Road, Durham DH1 3LE, UK}

\author[0000-0003-0931-0868]{Fatima Zaidouni}
\affiliation{MIT Kavli Institute for Astrophysics and Space Research, MIT, 77 Massachusetts Avenue, Cambridge, MA 02139, USA}

\begin{abstract}
X-ray reverberation mapping is a powerful technique for probing the innermost accretion disk, whereas continuum reverberation mapping in the UV, optical, and infrared (UVOIR) reveals reprocessing by the rest of the accretion disk and broad-line region (BLR). We present the time lags of Mrk~817 as a function of temporal frequency measured from 14~months of high-cadence monitoring from Swift and ground-based telescopes, in addition to an XMM-Newton observation, as part of the AGN~STORM~2 campaign. The XMM-Newton lags reveal the first detection of a soft lag in this source, consistent with reverberation from the innermost accretion flow. These results mark the first simultaneous measurement of X-ray reverberation and UVOIR disk reprocessing lags---effectively allowing us to map the entire accretion disk surrounding the black hole. Similar to previous continuum reverberation mapping campaigns, the UVOIR time lags arising at low temporal frequencies are longer than those expected from standard disk reprocessing by a factor of 2--3. The lags agree with the anticipated disk reverberation lags when isolating short-timescale variability, namely timescales shorter than the H$\beta$ lag. Modeling the lags requires additional reprocessing constrained at a radius consistent with the BLR size scale inferred from contemporaneous H$\beta$-lag measurements. When we divide the campaign light curves, the UVOIR lags show substantial variations, with longer lags measured when obscuration from an ionized outflow is greatest. We suggest that, when the obscurer is strongest, reprocessing by the BLR elongates the lags most significantly. As the wind weakens, the lags are dominated by shorter accretion disk lags.
\end{abstract}

\section{Introduction} \label{sec:intro}

Active Galactic Nuclei (AGN) are powered by accretion of material onto a supermassive black hole---a process that in turn releases enough energy in the form of electromagnetic radiation and mechanical outflows that is thought to play a critical role in the evolution of the host galaxy \citep[for a review, see][]{Fabian_2012}. We are unable to spatially resolve the innermost regions around the black hole, including the accretion disk and the broad line region (BLR), except for a few cases \citep[e.g.,][]{Russell_2018,Gravity_2018,Gravity_2020}. The central ionizing source irradiates these circumnuclear regions, which then respond after a time delay on the order of the light travel time to each emitting region. Reverberation mapping is a technique that allows us to constrain the spatial separation of circumnuclear regions by instead measuring these light travel times, or \textit{time lags} \citep[e.g.,][]{Blandford_1982, Peterson_2004, Bentz_2009, Fausnaugh_2016, Cackett_2018, Edelson_2019, Cackett_2020}. If we assume that the emission observed at each wavelength is dominated by a given region, then the light travel time between regions can be estimated from the time delays between variability in different wavebands. 

X-ray reverberation mapping probes the innermost accretion disk by measuring the time delays between X-ray bands; most commonly, between an energy band dominated by the continuum of the central X-ray corona and another dominated by the reflection \citep[i.e., the X-ray spectral signatures of reprocessing, including the soft excess below $\sim$2~keV and the Fe~K$\alpha$ line at 6.4~keV;][]{Ross_2005, Garcia_2010, Zoghbi_2011, DeMarco_2013, Uttley_2014, Kara_2016}. Whereas X-ray reverberation mapping accesses radii up to hundreds of gravitational radii ($R_g = GM/c^2$), continuum reverberation mapping in the ultraviolet, optical, and infrared (UVOIR) reveals reprocessing by the remainder of the accretion disk and the BLR, up to radii of $\sim10^5~R_g$ \citep[for a recent review, see][]{Cackett_2021}. In the canonical picture, coronal X-rays are thermally reprocessed by the disk (i.e., the disk is heated and then re-emits at longer wavelengths), producing correlated variability with a (light-crossing time) delay. The coronal X-rays reach the inner/hotter parts of the disk before reaching the outer/colder parts, thus producing longer time lags at longer wavelengths. To be specific, the lag amplitudes are expected to follow $\tau \propto \lambda^{4/3}$ when assuming the temperature profile of a standard \cite[e.g.,][]{SS_1973} accretion disk  \citep[][]{Collier_1998, Collier_1999, Cackett_2007}. The normalization of this relation depends on the mass and accretion rate of the black hole, in addition to physical properties of the disk \citep{Fausnaugh_2016}.

Recent campaigns using high-cadence observations from Swift and ground-based telescopes have been carried out for $\sim$10 AGN \citep[e.g.,][]{McHardy_2014, Shappee_2014, Edelson_2015, Fausnaugh_2016, Cackett_2018, McHardy_2018, Edelson_2019, Cackett_2020, HernandezSantisteban_2020, Kara_2021, Vincentelli_2021, Kara_2022}. In addition to these was AGN STORM, the large coordinated campaign for NGC~5548 \citep{DeRosa_2015}, which combined monitoring by Swift \citep{Edelson_2015} with spectroscopic monitoring by the Hubble Space Telescope (HST) and ground-based photometry \citep{Fausnaugh_2016} and spectroscopy \citep{Pei_2017}. Among many fascinating results was the discovery of the ``BLR holiday" \citep{Goad_2016}, a period 75 days after the start of the HST campaign in which the variations of the continuum and emission lines decorrelated. This decoupling was attributed to line-of-sight obscuration by a variable disk wind \citep{Dehghanian_2019a, Dehghanian_2019b}. 

These campaigns have expanded our catalog of continuum lag measurements tremendously, but have also solidified several mysteries from how the lags depart from theory. While the lags have been well described by the expected $\tau\propto\lambda^{4/3}$ relation, the normalizations of this relation have been larger than predicted, typically by a factor of 2--3 \citep[see, e.g.,][to name a few]{Edelson_2015, Fausnaugh_2016, Edelson_2019, Cackett_2020, Kara_2021}. The lags in the \textit{U} band near 3500~\AA\ are especially longer than predicted, exceeding even the best-fit $\tau\propto\lambda^{4/3}$ relation by roughly a factor of 2 \citep[see Figure~5 in][]{Edelson_2019}. Most importantly, the observed UV and X-ray variations have not been strongly correlated \citep[and are notably less correlated than that in the UV and optical, e.g.,][]{Starkey_2017, Edelson_2019, HernandezSantisteban_2020, Cackett_2020, Cackett_2023}, or, in some cases, completely uncorrelated \citep[e.g.,][]{Schimoia_2015, Buisson_2018}. This lack of correlation challenges our knowledge on the source of AGN disk heating and reverberation. These results are difficult to reconcile with the standard disk reprocessing picture in which the coronal X-rays are expected to generate the variability at longer wavelengths. Low X-ray-UV correlations have been interpreted, however, as a limitation of measuring the lags using the cross correlation function (CCF), which assumes a static configuration for the source, while the X-ray corona is most likely dynamic \citep{Panagiotou_2022a}.

An understanding of the ubiquity of X-ray-UV lag discrepancies in these campaigns arose from spectroscopic monitoring of NGC~4593 by HST  \citep{Cackett_2018}. The HST lags revealed that the \textit{U}-band excess measured in the Swift light curves was actually part of a previously unresolved discontinuity in the lags at the Balmer jump (3646~\AA). This discovery corroborated the theory that reprocessing by the BLR diffuse line and continuum is ``contaminating" the disk reprocessing lags; that is, the response of the BLR elongates the lags across all UVOIR bands, but most significantly at the Balmer and Paschen jumps \citep[][]{Korista_2001, Lawther_2018, Korista_2019, Netzer_2020, Netzer_2022}. Disk variability occurring on long timescales, such as temperature fluctuations that move radially inwards and/or outwards, may also play a role in elongating the lags \citep{Arevalo_2008, Arevalo_2009, Kelly_2009, Burke_2021}.

With the BLR positioned at larger radii than the disk (or, at least where reprocessing by the disk occurs most prominently), the BLR is expected to affect the lags on timescales longer than those of the disk. An approach that computes the lags separately for different timescales is therefore pivotal in order to distinguish reprocessing from the disk versus the BLR. Instead, the CCF method of \cite{Peterson_1998} has been used by a vast majority of the campaigns, which has been shown to be dominated by the variability on long timescales and thus reprocessing by the BLR \citep{Cackett_2022, Lewin_2023}. A common approach for removing the long-timescale contributions from the BLR is to first ``detrend" the light curves by subtracting from the time series a low-degree polynomial or a moving boxcar average \citep[e.g.,][]{McHardy_2018, HernandezSantisteban_2020, Pahari_2020, Vincentelli_2021, Lewin_2023, Cackett_2023}. In some cases \citep[e.g., NGC~4593 and Fairall~9;][]{McHardy_2018, HernandezSantisteban_2020}, applying this technique to the campaign light curves resolved the lag discrepancies, resulting in lags consistent with those expected from disk reprocessing.

The lags can also be computed as a function of temporal frequency\footnote{All mentions of ``frequency" refer to \textit{temporal frequency}, the inverse of which describes the timescale of the variability, as opposed to the frequency of light (wavelength or energy will always be used).} directly using Fourier techniques (hereafter, the \textit{frequency-resolved} lags). This approach isolates the variability on specific timescales, allowing for a more-straightforward analysis and robust modeling of the geometry required to reproduce the lags (using transfer functions). Initially, the CCF lags are generally consistent with the lags at low frequencies \citep{Cackett_2022, Lewin_2023}; after detrending the light curves, the CCF lags agree with higher frequency lags \citep[e.g., detrending with a low-degree polynomial in Mrk~335 accessed frequencies higher by a factor of 2--3;][]{Lewin_2023}. As a result, Fourier techniques have played a critical role in AGN timing analysis: isolating low frequencies reveals lags commonly attributed to propagating fluctuations in the mass accretion rate \citep{Lyubarskii_1997, Kotov_2001, Arevalo_2006, Yao_2023, Secunda_2023}, and, at high frequencies, the signatures of reprocessing (i.e., reverberation) by the innermost accretion flow \citep[e.g.,][]{Zoghbi_2011, DeMarco_2013, Kara_2016}. 

Computing the frequency-resolved lags, however, requires the light curves to be evenly sampled---a criterion satisfied more commonly by X-ray light curves. Several methods have been developed to compute the frequency-resolved lags from unevenly sampled light curves. Briefly, the maximum likelihood technique of \cite{Miller_2010} and \cite{Zoghbi_2013} consists of fitting a model for the power spectral density (PSD), which is then used to compute the cross-spectrum and thus the frequency-resolved lags. Another method is modeling the variability in each band using Gaussian processes (GPs), from which one can then draw evenly sampled realizations of the light curves used to constrain the frequency-resolved lags. For context, GPs have been applied in machine learning research extensively for decades, especially after  \cite{Neal_1995} demonstrated that infinitely complex Bayesian neural networks converge to GPs. In the astrophysics community, GPs have shown success for modeling light curves of asteroids \citep{Lindberg_2021}, stars \citep{Brewer_2009, Czekala_2017, Nicholson_2022}, and AGN \citep{Kelly_2014, 2018MNRAS.475.2051K, Wilkins_2019, Griffiths_2021, Lewin_2022, Lewin_2023}, and for generative modeling \citep[e.g., quasar spectra;][]{Eilers_2022}. When modeling AGN variability, GPs have been found to preserve both the underlying autocorrelation functions, and thus the PSD \citep[][]{Wilkins_2019, Griffiths_2021}, and the phase information between light curves. Indeed, time lags have been recovered within a fractional error of a few percent using simulations and real data \citep[][]{Wilkins_2019, Lewin_2022, Lewin_2023}. 

Frequency-resolved lags have been computed for three reverberation mapping campaigns so far: NGC~5548 and Fairall~9 \citep[both using the maximum likelihood technique;][]{Cackett_2022, Yao_2023} and Mrk~335 \citep[using GPs;][]{Lewin_2023}. In all sources, the lags at low frequencies exceed those expected from disk reprocessing by a factor of 2--3. At higher frequencies, the lags decrease in amplitude, and agree with disk reprocessing when homing in on timescales shorter than the BLR radius inferred from the H$\beta$ lag \citep{Cackett_2022, Lewin_2023}. The \textit{U}-band lag excesses were also absent at these high frequencies, further corroborating the BLR as a leading culprit causing the lag discrepancies. In both cases, simple disk models could not reproduce the long low-frequency lags. Instead, an additional model component was required to account for reprocessing from beyond the disk, namely at a radius consistent with that of the BLR. 

AGN~STORM~2 is the next large multiwavelength reverberation mapping campaign, featuring 15 months of high-cadence monitoring of the nearby Seyfert~1 galaxy Mrk~817 ($z$ = 0.031455) by HST, Swift, NICER, and ground-based telescopes, with spectroscopy and photometry in the optical and near-IR presented in \cite{Montano_2023}. Several simultaneous observations were also carried out by XMM-Newton and NuSTAR. \cite{Kara_2021}, hereafter Paper~I, presented the results of the first three months of the campaign, including the unprecedentedly low X-ray flux due to obscuration. \cite{Homayouni_2023}, hereafter Paper~II, presented the HST observations, including UV emission line reverberation results. \cite{Partington_2023}, hereafter Paper~III, presented the NICER observations, with spectral analysis revealing that the variable observed X-ray flux is coincident with changes in the column density and ionization of the aforementioned obscurer. \cite{Cackett_2023}, hereafter Paper~IV, presented the Swift observations with a focus on the UV/optical continuum variability and reverberation. \cite{Homayouni_2023_2}, hereafter Paper~V, further studied the varying response of the broad UV emission lines found in Paper~II. The variability of the disk is studied in Paper~VI \citep{Neustadt_2024} using temperature fluctuation maps resolved both in time and radius. \cite{Zaidouni_2024}, hereafter Paper~IX, presented the XMM-Newton and NuSTAR observations, including analysis of high-resolution grating spectra revealing that the obscurer is a multi-phase disk wind.

Here in Paper~VII, we present frequency-resolved timing analysis of the XMM-Newton, Swift, and ground-based light curves, in effect mapping the innermost accretion disk (X-ray reverberation mapping), out to the outermost disk and the BLR (UVOIR continuum reverberation mapping). In Section~\ref{sec:obs}, we introduce the observations and data reduction. In Section~\ref{sec:methods}, we detail the methods, namely Fourier analysis and GP regression, used to produce the results presented in Section~\ref{sec:results}. The results are modeled in Section~\ref{sec:modeling} and discussed in Section~\ref{sec:discuss}. 

\section{Observations} \label{sec:obs}

\subsection{Swift + Ground-based Campaign}
\begin{figure}[t!]
    \centering
    \includegraphics[width=0.92\columnwidth]{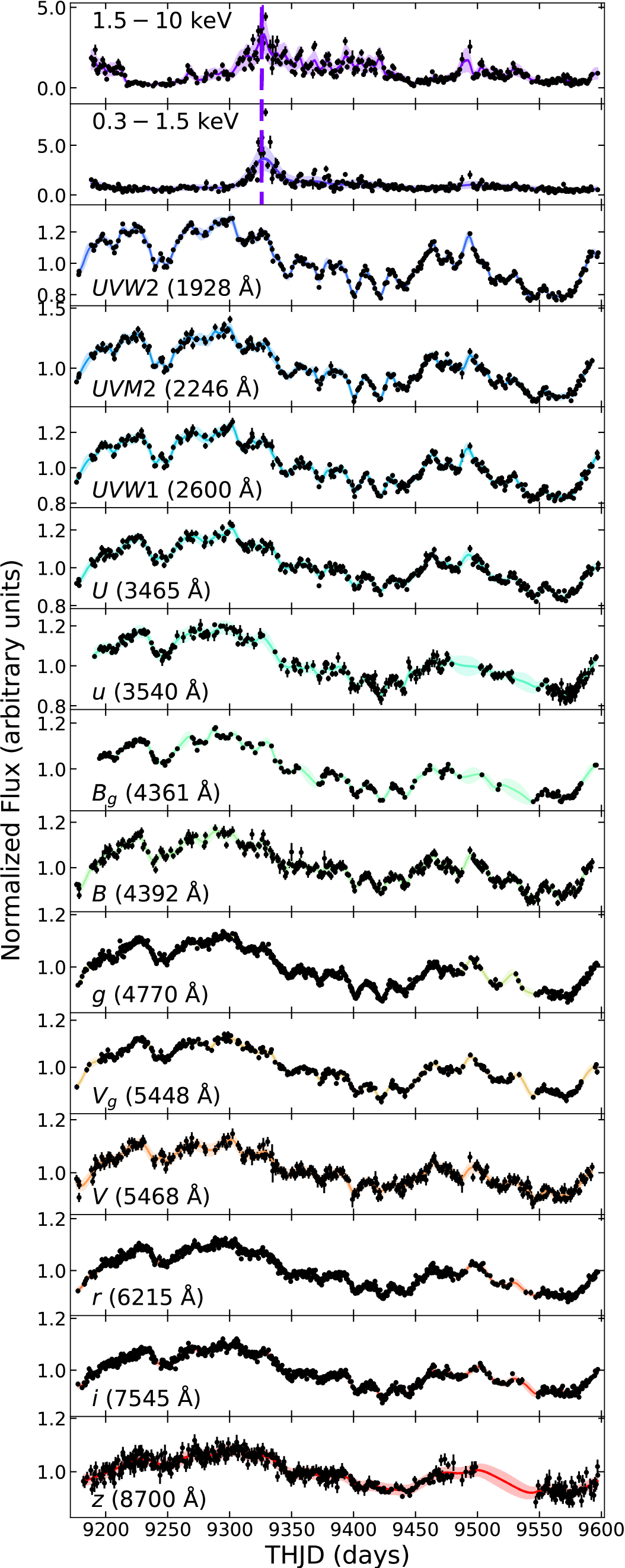}
    \caption{AGN STORM 2 light curves from Swift and ground-based telescopes from THJD = 9177--9597. The ground-based filters are lowercase, except for the ground-based Johnson/Bessel $B_g,V_g$. A dashed vertical line in the X-ray bands indicates the XMM-Newton observation used in our analysis. The average of 1000 GP realizations is shown by colored lines, with 1$\sigma$ shaded regions.}
    \label{fig:ibrm_lcs}
\end{figure}
The Neil Gehrels Swift Observatory \citep{Gehrels_2004} performed daily monitoring of Mrk~817 for 15 months (2020 November 22 through 2022 February 24), and these Swift AGN STORM 2 results were initially presented in Paper~IV \citep{Cackett_2023}. Our analysis focuses on data collected simultaneously by Swift and ground-based telescopes in the 420-day time window THJD=9177--9597 shown in Figure~\ref{fig:ibrm_lcs}, where the Truncated HJD (THJD) is defined as THJD=HJD--2450000. This time range is roughly 90 per cent of that used by \cite{Cackett_2023}, since we exclude the last $\sim$40 days of the campaign to remove a large gap in the Swift data resulting from the observatory entering safe mode due to the failure of a reaction wheel. Interpolating over this gap using GP regression results in larger uncertainties on the lags, although the results are consistent with those shown here. The Swift X-ray light curves were produced using the Swift-XRT data products generator\footnote{ \url{https://www.swift.ac.uk/user_objects/index.php}} \citep{Evans_2007, Evans_2009}. The background-subtracted count rates were produced with per-observation binning in a soft band (0.3--1.5~keV) and a hard band (1.5--10~keV). We refer the reader to \cite{Cackett_2023} for details on the Swift UVOT data reduction. 

The ground-based observations were carried out by the following observatories: Las Cumbres Observatory Global Telescope network \citep{2013PASP..125.1031B}, Liverpool Telescope \citep{10.1117/12.551456}, the Calar Alto Observatory, Wise Observatory \citep{2008Ap&SS.314..163B}, the Yunnan Observatory, and the Zowada Observatory \citep{2022PASP..134d5002C}. The intercalibrated ground-based photometry is presented in \cite{Montano_2023}. The SDSS $u^\prime g^\prime r^\prime i^\prime z^\prime$ and Pan-STARRS $z_s$ filters were used, with the measurements in both the SDSS and Pan-STARRS \textit{z}-labeled filters combined. We hereafter refer to the filters as the \textit{ugriz} bands. The ground-based Johnson/Bessel $B,V$ filters are denoted $B_g, V_g$ to avoid confusion with the Swift filters. We refer the reader to \cite{Kara_2021} for details on the ground-based data reduction, and a full description of the ground-based campaign and photometry will be presented by \cite{Montano_2023}.

\subsection{XMM-Newton}
\begin{figure}[t!]
    \centering
    \includegraphics[width=\columnwidth]{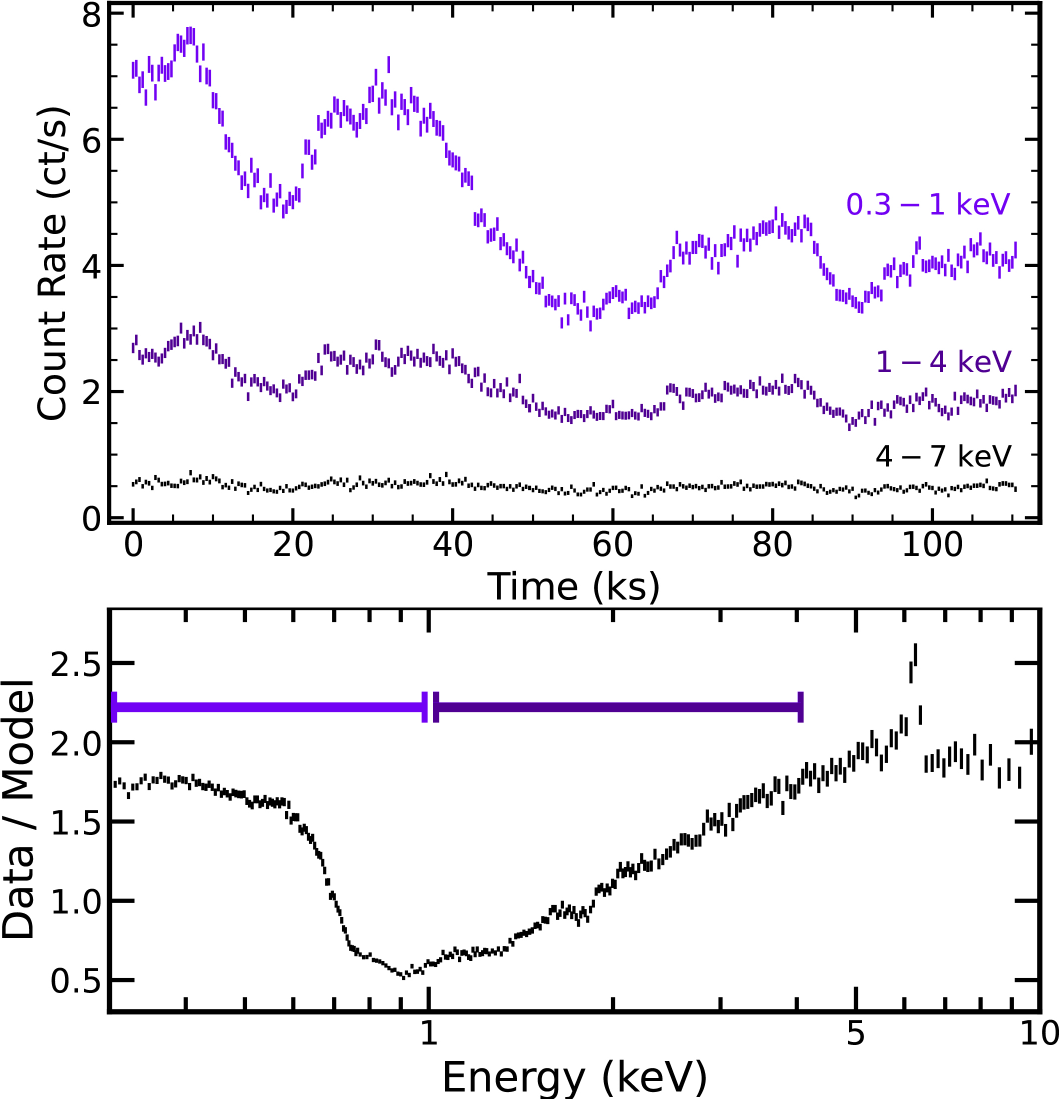}
    \caption{\textit{Top:} Light curves in three energy bands for the XMM-Newton observation used in our analysis. The lower energy bands spanning 0.3--4~keV exhibit clear variability, from which we compute frequency-resolved time lags probing the innermost accretion flow. \textit{Bottom:} The XMM-Newton spectrum from the same observation divided by a power-law model with a photon index of 2. Purple horizontal lines visualize the 0.3--1~keV and 1--4~keV energy ranges used in the timing analysis to isolate reflection and the direct coronal continuum, respectively.}
    \label{fig:xmm_lcs}
\end{figure}

XMM-Newton carried out a total of four observations throughout the AGN STORM 2 campaign, which were presented in Paper~IX \citep{Zaidouni_2024}. For all but one observation, the X-ray flux was an order of magnitude fainter than historical observations due to obscuring gas along the line of sight. It was serendipitous that the 2021 April observation (obsid:~0882340601) coincided with the only prominent X-ray flare observed in the campaign on THJD = 9329, as shown in Figure~\ref{fig:ibrm_lcs}. The high-resolution grating spectra of this observation showed strong narrow absorption lines, from which \cite{Zaidouni_2024} revealed that the obscuring gas is a multi-phase, ionized wind. The observation also provides a fortuitous opportunity for timing analysis, with high-amplitude variability exhibited by the light curves in Figure~\ref{fig:xmm_lcs}.

We performed the same standard data reduction procedure as \cite{Zaidouni_2024}: the EPIC-pn data were reduced using the XMM-Newton Science Analysis System (\textsc{sas v. 19.0.0}). The light curves were extracted using circular source and background regions, both with 35 arcsec radii, and then binned to 30-second bins. Both regions are located on the same chip and not near the edges of the chip. We avoided background flares by constructing a good time interval filter with a background count rate cutoff of 0.4 cts s$^{-1}$, in addition to avoiding spurious detections using the conditions ``PATTERN$\leq$4" and ``FLAG$==$0." Soft photon background flares were also cleaned from both ends of the observation, resulting in a final exposure of $\sim$120~ks. 

\section{Methods} \label{sec:methods}

We aim to measure how the variability operating on different timescales/frequencies in each waveband lags or leads that of another. Computing these \textit{frequency-resolved lags} using Fourier techniques requires the data to be evenly sampled (i.e., without gaps). While this criterion is satisfied in the case of the XMM-Newton observation, the sampling rate of the Swift and ground-based telescope data constantly changes over the course of the campaign. This obstacle is commonly overcome by either maximizing the likelihood of a model for 1) the PSD directly \citep[e.g.,][]{Zoghbi_2013} or 2) the observed variability to infer data in the gaps (i.e., regression). 

Here, we choose the latter by modeling the observed variability in each waveband using GPs, allowing us to draw equally sampled light curve realizations from which we compute the frequency-resolved lags. Since the variability differs across wavebands, a unique GP is trained in each band in a self-contained manner (independent of the data in the other bands). The conditional posterior from which the realizations are drawn is informed by the empirical variability operating on all timescales present in the light curve. The realizations should thus reflect the underlying variability process, as previously demonstrated via the faithful recovery of underlying AGN autocorrelation functions when averaging across realizations \citep{Griffiths_2021}. Previous work has also shown that GP regression preserves the phase information (i.e., time lags) between light curves with data quality similar to that of our Swift and ground-based data \citep[e.g.,][]{Wilkins_2019, Lewin_2023}. We nonetheless demonstrate the successful recovery of simulated time lags given our specific data quality below in Appendix~\ref{sec:sims}, where we present the effects of GP regression on the lags and their uncertainties for the data quality in all wavebands.

\subsection{Fourier-resolved timing analysis} \label{subsec:fourier}

We compute the lags as a function of frequency using a standard procedure involving Fourier techniques \citep[for a review, see][]{Uttley_2014}. For the regularly sampled XMM-Newton observation, we can perform the procedure immediately: the cross-spectrum is computed between a light curve dominated by reflection (0.3--1~keV) and another dominated by the direct continuum of the corona (1--4~keV). These energy bands were selected based on the spectral features shown in Figure~\ref{fig:xmm_lcs}, namely the soft excess commonly attributed to reprocessing/reflection below 1~keV \citep[e.g.,][]{Ross_2005, Garcia_2014} and a featureless power law from 1--4~keV. This selection is in agreement with modeling of the spectral energy distribution (SED) in Paper~IX \citep{Zaidouni_2024}. The cross-spectrum is then binned into coarser frequency bins, each centered at frequency $\nu_i$. The phase of the cross-spectrum at each frequency is converted to a final time lag by dividing by $2\pi\nu_i$, producing a \textit{lag-frequency spectrum}. 

We also measure how 9 finer energy bands lag or lead a common reference band---in this case, the 0.3--10~keV broadband---in a given frequency range (i.e., a \textit{lag-energy spectrum}). In this case, the light curve in each energy band of interest is subtracted from that of the reference band to remove Poisson noise that would otherwise be correlated between light curves. The lag-energy spectrum reveals how each energy band contributes to the lags, and thus is useful for studying the causal relationship between different spectral components \citep{Uttley_2014}. 

In order to compute the frequency-resolved lags for the irregularly sampled Swift and ground-based observations, we instead apply the aforementioned procedure to GP realizations of the light curve. As introduced in Section~\ref{subsec:gps}, these realizations have been informed by the empirical variability at all timescales present in the data. A lag-frequency spectrum is then computed for each of the 1000 realization pairs (1000 realizations in each waveband of interest and in the \textit{UVW2} reference band). The final lag-frequency spectrum and $1\sigma$ uncertainties are produced by computing the mean and standard deviation of this distribution of 1000 lag-frequency spectra. This number of realizations was selected based on convergence of the lag distribution: a two-sample C-vM test between the empirical cumulative distribution function (ECDF) of the lags produced from the first 1000 samples to that from 5000 samples results in a \textit{p}-value of $\sim$0.3, and we are thus unable to reject the null hypothesis that both distributions are the same.

The minimum frequency accessible using this approach is limited by the length of the light curve ($\nu_{\text{min}} = 1/L$ for length $L$), whereas the maximum frequency is set by the sampling rate ($\nu_{\text{max}} = 1/(2\Delta t)$ for sampling rate $\Delta t$). The XMM-Newton observation allows us to probe frequencies on the order of $10^{-5}-10^{-3}$ Hz, equivalent to studying radii from $\sim 1- 100~R_g$, i.e., the innermost accretion disk. The Swift and ground-based data allow us to study variability operating on much longer timescales, down to frequencies on the order of $10^{-3}~\text{day}^{-1}$, equivalent to radii from $\sim 100 - 10^5~R_g$, thus probing the outermost parts of the accretion disk and the BLR.

\subsection{Gaussian process regression} \label{subsec:gps}

We include a brief overview of GP regression here, but refer the reader to more detailed introductions by \cite{Rasmussen_2006,Wilkins_2019, Griffiths_2021}.

We can consider our light curve data to be a vector of fluxes $\textbf{d}$ observed at times $\textbf{t}$. The ``Gaussian" in ``Gaussian process" arises in that we assume the data have been drawn from a multivariate Gaussian distribution; or, in other words, that the observed data are a realization drawn from a GP posterior. This procedure assumes the data are normally (or log normally) distributed, the validity of which we explore in Appendix~\ref{sec:normal}, where we conclude that the data in all wavebands agree better with a log-normal distribution and thus train the GPs on the log-transformed flux values. The properties of this multivariate Gaussian is informed by the data, including the \textit{mean function} $m(t)=\mathbb{E}[f(t)]$ ($\mathbb{E}[x]$ denotes the expected value of $x$ and $f(t)$ the function of fluxes observed at time $t_i$), and the covariance function, hereafter referred to as the \textit{kernel function}, $k(t,t’)=\mathbb{E}[(f(t) - m(t))(f(t^\prime) - m(t^\prime))]$. 

The mean function is taken to be $m=0$, as it is common practice to first standardize the data (subtract the mean of the light curve before dividing by the standard deviation). The kernel function encodes how the light curve deviates from the mean and thus models the observed variability. Functional forms for the kernel function commonly depend on the separation in time between points (i.e., $k(t,t’)=k(t-t')$), thus modeling the empirical variability on every timescale present. The choice of kernel function form has been found to impact the significance of lag recovery, depending on the data quality \citep{Griffiths_2021,Lewin_2022}. We detail the selection of the functional form of the kernel function in Appendix~\ref{sec:kernels}, choosing between kernel functions that have shown success in modeling AGN variability for a wide range of data quality \citep[e.g.,][]{Wilkins_2019, Griffiths_2021, Lewin_2022, Lewin_2023}. In summary, the rational quadratic kernel performs the best statistically in all bands, although the final lags are consistent across the three kernel forms. At all frequencies, the lags agree within 5\% of each other on average, with the lag uncertainties agreeing within 10\%. Every functional form comes with its own hyperparameters ($\theta$), each encoding a property of the variability, for instance timescales and amplitudes. These hyperparameters are determined by maximizing the likelihood of the model given the observed data; or, more commonly, by minimizing the negative log marginal likelihood \citep[NLML; Equation~17 in][]{Griffiths_2021}. 

Equally sampled realizations with flux values $\textbf{d}_*$ can then be generated by taking random draws from the conditional distribution $(\textbf{d}_*|\textbf{d})$ \citep[Equation~5 in][]{Wilkins_2019}, which is defined by the optimized Gaussian and the observed flux vector \textbf{d}. 

The architecture used for GP model training and subsequent regression was created by using \texttt{Scikit-learn}\footnote{\href{https://scikit-learn.org/}{https://scikit-learn.org/}} and the X-ray timing analysis package \texttt{pyLag}\footnote{\href{http://github.com/wilkinsdr/pylag}{http://github.com/wilkinsdr/pylag}} \citep{Wilkins_2019}. 

\section{Results} \label{sec:results}
\subsection{Innermost disk: X-ray reverberation mapping}

\begin{figure*}[t!]
    \centering
    \includegraphics[width=0.72\textwidth]{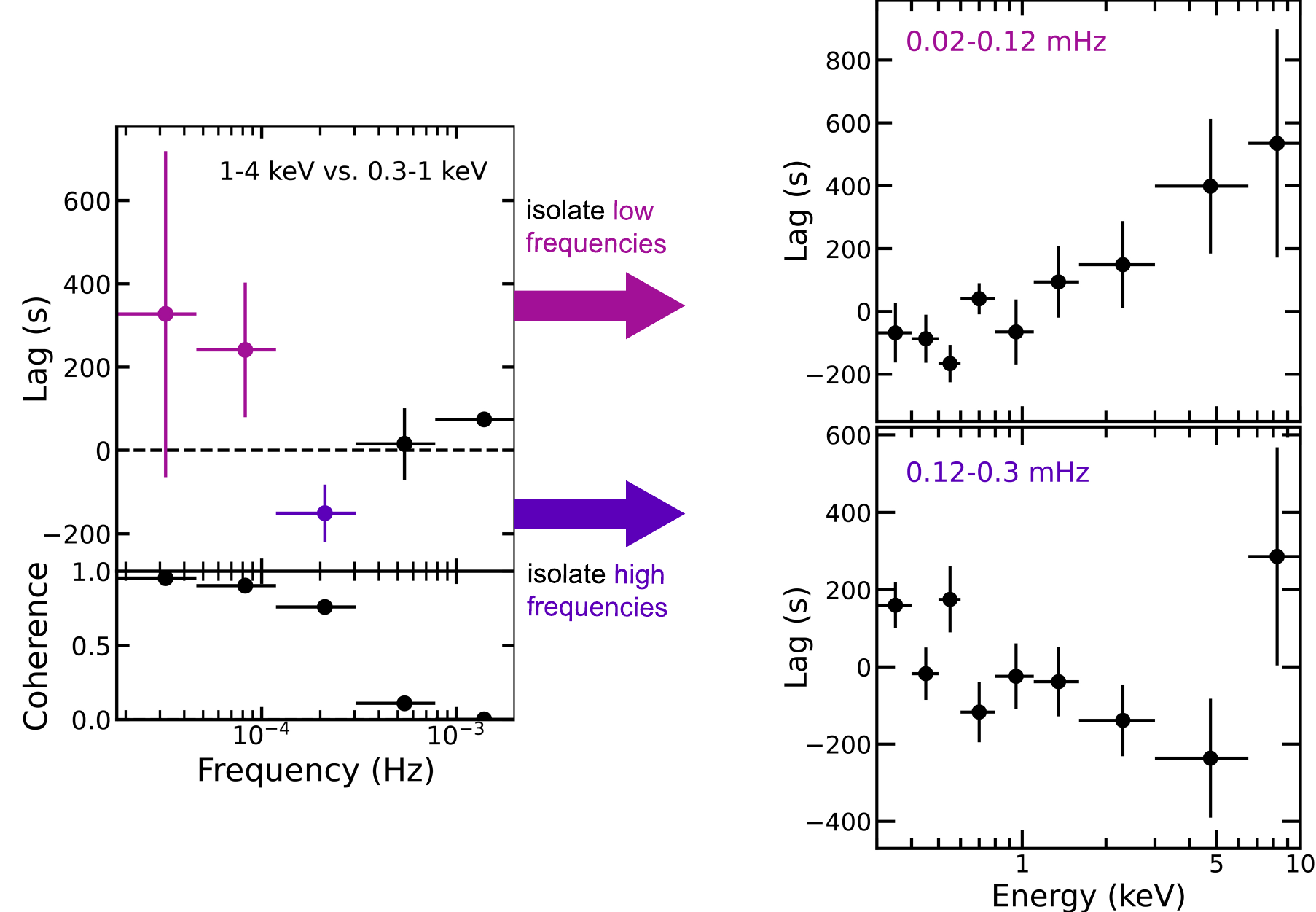}
    \caption{Key X-ray XMM-Newton time lag results: \textit{Left:} Lag-frequency spectrum with bias-subtracted coherences computed between a reflection-dominated band (0.3--1~keV) and a continuum-dominated band (1--4~keV). A positive lag indicates the hard band lagging the soft band. \textit{Right:} The lag-energy spectrum when isolating low frequencies (\textit{upper}) reveals lags increasing with energy, commonly attributed to the inward propagation of mass-accretion rate fluctuations. The lag-energy spectrum at high frequencies (\textit{lower}) instead exhibits features consistent with reverberation (albeit at low significance due to larger uncertainties above 3~keV), including peaks in the lags below 1~keV where we expect the soft excess, and in the 6--10~keV bin where we expect the Fe~K.}
    \label{fig:xray_reverb}
\end{figure*}

Due to heavy obscuration studied throughout the campaign in Paper~III \citep{Partington_2023}, the average X-ray flux is an order of magnitude lower than that observed at the time of the flare simultaneous with the XMM-Newton observation. The flare revealed prominent X-ray variability, as shown by the light curves in Figure~\ref{fig:xmm_lcs}), providing a momentary glimpse of the innermost accretion disk that we aimed to study with X-ray timing analysis.

Here, we present the frequency-resolved X-ray time lags of Mrk~817 computed from the $\sim$120-ks XMM-Newton observation using the procedure described in the previous section. The lags span temporal frequencies of $10^{-5}-10^{-3}$~Hz and were computed between an energy band dominated by the coronal continuum (1--4~keV, hereafter the hard band) and another band dominated by the reprocessed emission (0.3--1~keV, hereafter the soft band). 

The resulting lag-frequency spectrum and associated (bias-subtracted) coherence values are shown on the left side of Figure~\ref{fig:xray_reverb}. The lag-frequency spectrum exhibits a shape typical for AGN \citep[e.g.,][]{Zogbhi_2010, Kara_2013, DeMarco_2013}: at low frequencies, the hard band lags the soft band, whereas at slightly higher frequencies, the soft band instead lags the hard band (i.e., a \textit{soft lag}, $150\pm68~\text{s}$ in the $(1-3)\times10^{-4}~\text{Hz}$ bin). This is the first detection of a soft lag in this source, as the only archival XMM-Newton observation of Mrk~817 (previous to the STORM 2 campaign) was 11-ks long \citep{Winter_2011}. Both the measured lag amplitude and frequency of this soft lag are consistent with the lag-mass scaling relations found by \cite{DeMarco_2013}, which predict a lag amplitude of $210\pm80$~s at $(1.6\pm0.3)\times10^{-4}~\text{Hz}$ for a black hole mass of $M_{\text{BH}} = 3.85\times 10^7~\text{M}_\odot$. The hard and soft lags have high coherence ($>0.90$ for the hard lag and $>0.75$ for the soft lag), indicating a high degree of linear correlation between light curves.

In both frequency ranges at which we detect a hard or soft lag, we computed a lag-energy spectrum by measuring how the variability in finer energy bands lags or leads a common (0.3--10~keV) reference band. The lag-energy spectrum at low frequencies (the top-right panel of Figure~\ref{fig:xray_reverb}) increases roughly monotonically with energy. This common shape in the lag-energy spectrum \citep[e.g.,][]{Kara_2016} is often attributed to fluctuations in the mass-accretion rate that propagate inwards from colder to hotter regions, thus resulting in soft photons that respond before the hard photons \citep{Lyubarskii_1997, Kotov_2001, Arevalo_2006}. The lag-energy spectrum when isolating the higher frequencies where we detected a soft lag (the bottom-right panel of Figure~\ref{fig:xray_reverb}) instead shows an anticorrelation between lag and energy, except for the 6--10~keV bin. The lag-energy spectrum thus appears similar to previous AGN lag spectra exhibiting reverberation signatures, including peaks in the lags below 1~keV where we expect the so-called soft excess, and in the 6--10~keV bin where we expect the Fe~K. We measure a $\sim$500-second difference in the lag between the 6--10~keV and 3--6~keV bins, albeit with large uncertainties due to low signal-to-noise ratio. This difference is consistent with the expected amplitude of the Fe~K lag based on the lag-mass scaling relation confirmed by \cite{Kara_2016}.

\subsection{Outermost disk/BLR: UVOIR continuum reverberation mapping}

Here, we present the frequency-resolved UVOIR time lags of Mrk~817 computed from the AGN STORM 2 Swift and ground-based campaign light curves shown in Figure~\ref{fig:ibrm_lcs}. Given the length of the campaign, we computed the lags using two treatments of the light curves: 1) using the full 420-day light curves and 2) when splitting the light curves into three epochs of equal length (140~days). These epochs fortuitously coincide with periods of high (Epochs~1 and 3) and low (Epoch~2) average column density in the obscurer, according to the NICER spectral analysis in Paper~III \citep{Partington_2023}. The Swift X-ray-UV lags were also measured, but a tentative lag (with coherence less than 0.2) is measured in only the second epoch as a result of the X-ray flare. As such, the X-ray-UV lags are only discussed when presenting the epoch-resolved lags (Section~\ref{subsubsec:epoch_lags}). 

\subsubsection{Full campaign lags} \label{subsubsec:campaign_lags}

\begin{figure*}[t!]
    \centering
    \includegraphics[width=\textwidth]{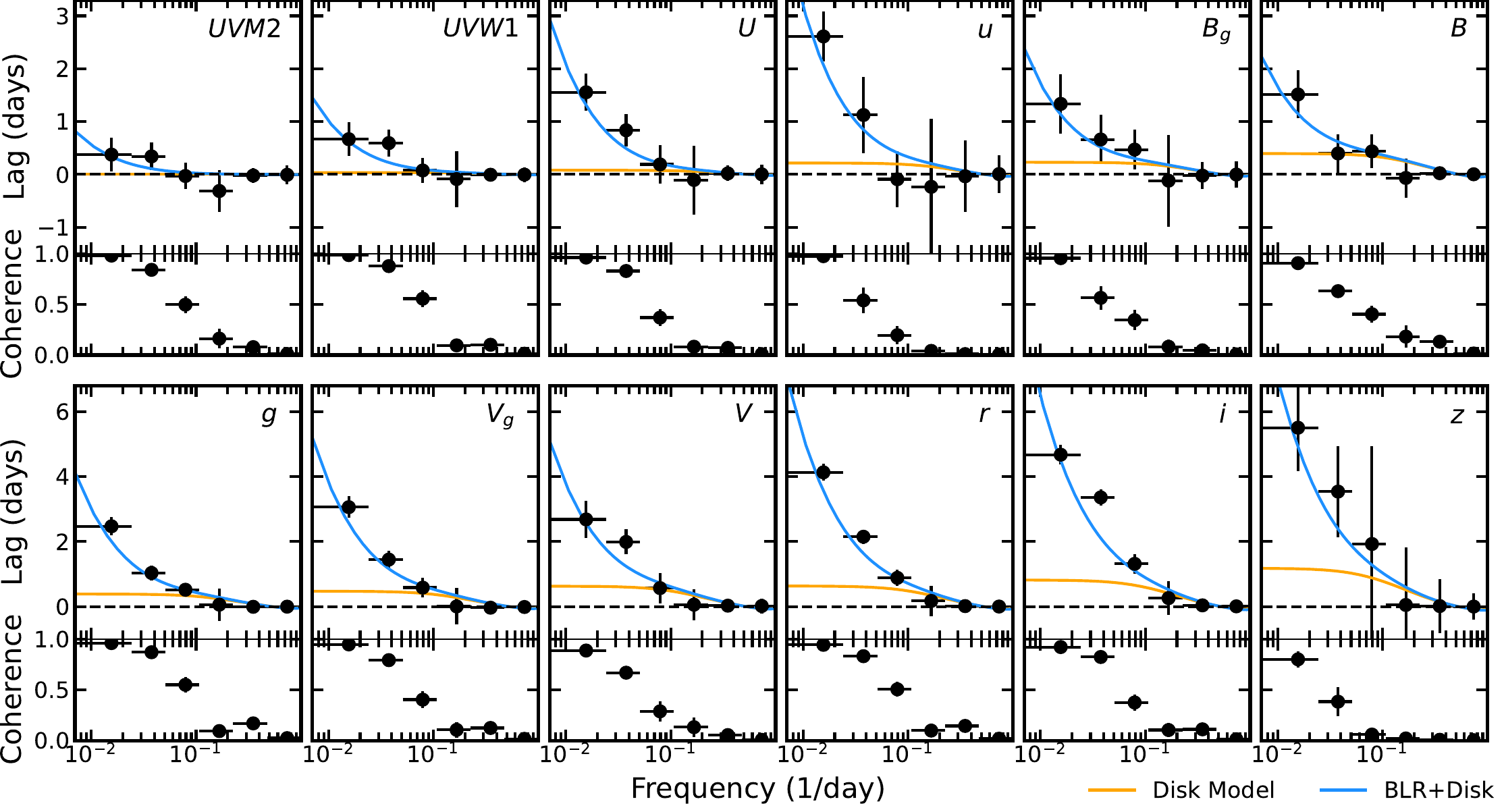}
    \caption{Lags as a function of frequency in the UVOIR bands (with respect to the \textit{UVW2} reference band), with corresponding bias-corrected coherence values below. The accessible frequencies  ($\sim10^{-2} - 10^{0}~\text{day}^{-1}$, i.e., $\sim10^{-8} - 10^{-6}~\text{Hz}$), are over two orders of magnitude lower than those studied using the XMM-Newton observation. The lags predicted by the disk reprocessing model (orange) given the mass and accretion rate of Mrk~817 poorly describe the measured lags at low frequencies. The fit improves significantly when we include a simple model component to account for additional contribution to the lags from a distant reprocessor (final model in blue). The radius of this reprocessor is constrained at $\sim$23 days, consistent with previous measurements of the H$\beta$ lag indicative of the BLR size scale.}
    \label{fig:lfs_full}
\end{figure*}
\begin{figure*}[t!]
    \centering
    \includegraphics[width=0.9\textwidth]{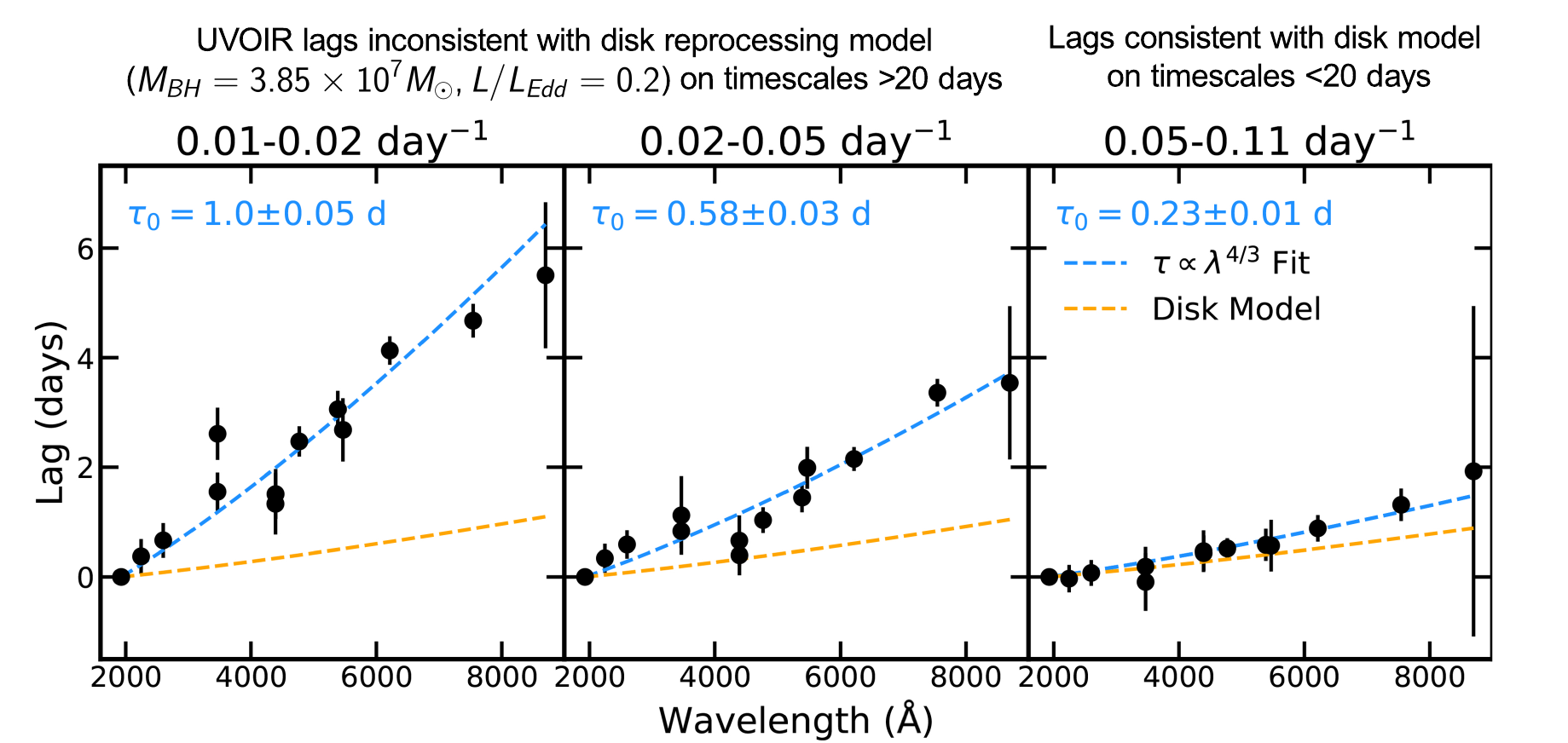}
    \caption{UVOIR lags as a function of wavelength in the lowest three frequency bins. The lags are fit with a $\tau \propto \lambda^{4/3}$ relation (dashed blue line), with best-fit normalization values ($\tau_0$) shown. We compare the lags to the expected lag-wavelength relation given the mass and accretion rate of Mrk~817 using the disk reprocessing model in each frequency range (dashed orange line). At higher frequencies, the lags approach the expected lag-wavelength, with the lags in the $0.05-0.11~\text{day}^{-1}$ range (right-most panel) roughly consistent with the disk reprocessing model. If this discrepancy is due to additional reprocessing from beyond the disk, it is occurring on timescales longer than 20~days, which is consistent with the BLR size scale inferred from the 23-day H$\beta$ lag.}
    \label{fig:lws_full}
\end{figure*}

The lags and bias-corrected coherence were computed with respect to the \textit{UVW2} band centered at 1928~\AA\ for the full 420-day Swift and ground-based light curves in six bins spanning $7\times10^{-3}-10^0~\text{day}^{-1}$. This frequency range translates to light travel times from the black hole out to roughly $10^5~R_g$, thus probing reprocessing by the outermost accretion disk and the BLR \citep{Cackett_2021}. The lowest frequency we probe is set by the length of the observation, but we expect to see contributions from emission by the dusty torus on these long timescales \cite{Netzer_2022}, which may complicate the analysis.

The lags as a function of frequency for all of the wavebands are shown in Figure~\ref{fig:lfs_full}, alongside lag models detailed in the following section. These lags are also provided in the Appendix (Table~\ref{tab: full_lags}). Both lags and their coherence values decrease with frequency, similar to those reported for NGC~5548 \citep{Cackett_2022} and Mrk~335 \citep{Lewin_2023}. The lags continue to increase at lower frequencies (i.e., the lags do not level out), indicating additional reprocessing beyond timescales of 20 days. This is consistent with reprocessing by the BLR based on the 23-day H$\beta$ lag measured in Paper~I \citep{Kara_2021}. 

The average coherence measured in the two lowest frequency bins is high  at $>0.9$ and $>0.75$, respectively. The coherence is notably lower---producing larger uncertainties on the lags---in the ground-based \textit{u, $B_g$, z} bands, presumably due to coarse data sampling near the end of the campaign. We nonetheless verified that simulated lags are faithfully recovered in these cases where GP regression over substantial gaps produces substantive drops in coherence, as detailed in the Appendix. The drop in coherence on short timescales (even without the use of GPs) suggests that white noise dominating the variability at higher frequencies may be breaking the correlation between light curves, although the washing out of variability in the disk may also be playing a role.

The lags in the lowest three frequency bins are presented as a function of wavelength in Figure~\ref{fig:lws_full}. Each set of lags is generally well described by a $\tau \propto \lambda^{4/3}$ relation, as expected from reprocessing by a standard accretion disk \citep{Cackett_2007}.  In detail, the lags in each frequency range were independently fit with the function $\tau = \tau_0 [(\lambda/\lambda_0)^{4/3} - 1]$, fitting for the normalization ($\tau_0$) given the reference band rest-frame wavelength for the \textit{UVW2} band ($\lambda_0=1869$~\AA). This results in the following best-fit normalization values for the three lowest frequency bins (in order of increasing frequency): $\tau_0 = 1.00\pm0.05, 0.58\pm0.03, 0.23\pm0.01~\text{days}$. We also note a discrepancy between the lags in the \textit{U} and \textit{u} bands: in the lowest frequency bin, the ground-based \textit{u}-band lag is $>1\sigma$ larger than that in the Swift \textit{U} band. A very similar discrepancy was found in NGC~5548 \citep{Cackett_2022}. 

As a point of comparison, we modeled the frequency-resolved lags expected from disk reprocessing given the mass and accretion rate of Mrk~817 by generating impulse-response functions, as detailed in Section~\ref{sec:modeling}. The lags at low frequencies ($<0.05~\text{day}^{-1}$, corresponding to timescales $>20~\text{days}$; see the two left-most panels in Figure~\ref{fig:lws_full}) are significantly longer than those predicted by the disk reprocessing model--on average, by over a factor of 3 in the lowest frequency bin and a factor of 2 in the second-lowest frequency bin. This discrepancy is the most significant in the \textit{u} band (3465~\AA), where the lag exceeds the $\tau\propto\lambda^{4/3}$ fit by a factor of 2 in the lowest frequency bin, as also found in the CCF lags in Paper~IV \citep{Cackett_2023}. Similar levels of disagreement from disk reprocessing have been reported by previous continuum reverberation mapping campaigns, including especially long lags in the \textit{U} band  \citep[e.g.,][]{Cackett_2018, Edelson_2019, HernandezSantisteban_2020, Vincentelli_2021, Kara_2022}.

At higher frequencies ( $>0.05~\text{day}^{-1}$, corresponding to timescales $<20~\text{days}$; see the right-most panel in Figure~\ref{fig:lws_full}), the lags have decreased in size and show general agreement with the disk reprocessing model. The \textit{U}-band excess is also absent at these frequencies, with the lag now within 1$\sigma$ from the $\tau\propto\lambda^{4/3}$ fit. Given the H$\beta$ lag measured in Paper~I \citep[$\tau_{\text{H}\beta}=23.2\pm1.6~\text{days}$;][]{Kara_2021}, these findings are similar those in Mrk~335 and NGC~5548: the lags are consistent with disk reprocessing (including the \textit{U}-band excess) when homing in on timescales shorter than the H$\beta$ lag indicative of the BLR size scale. These results thus further support that reprocessing by the BLR is producing the longer-than-predicted lags reported by previous campaigns, as discussed in Section~\ref{sec:discuss}. 

\subsubsection{Epoch-resolved lags} \label{subsubsec:epoch_lags}
\begin{figure*}[t!]
    \centering
    \includegraphics[width=\textwidth]{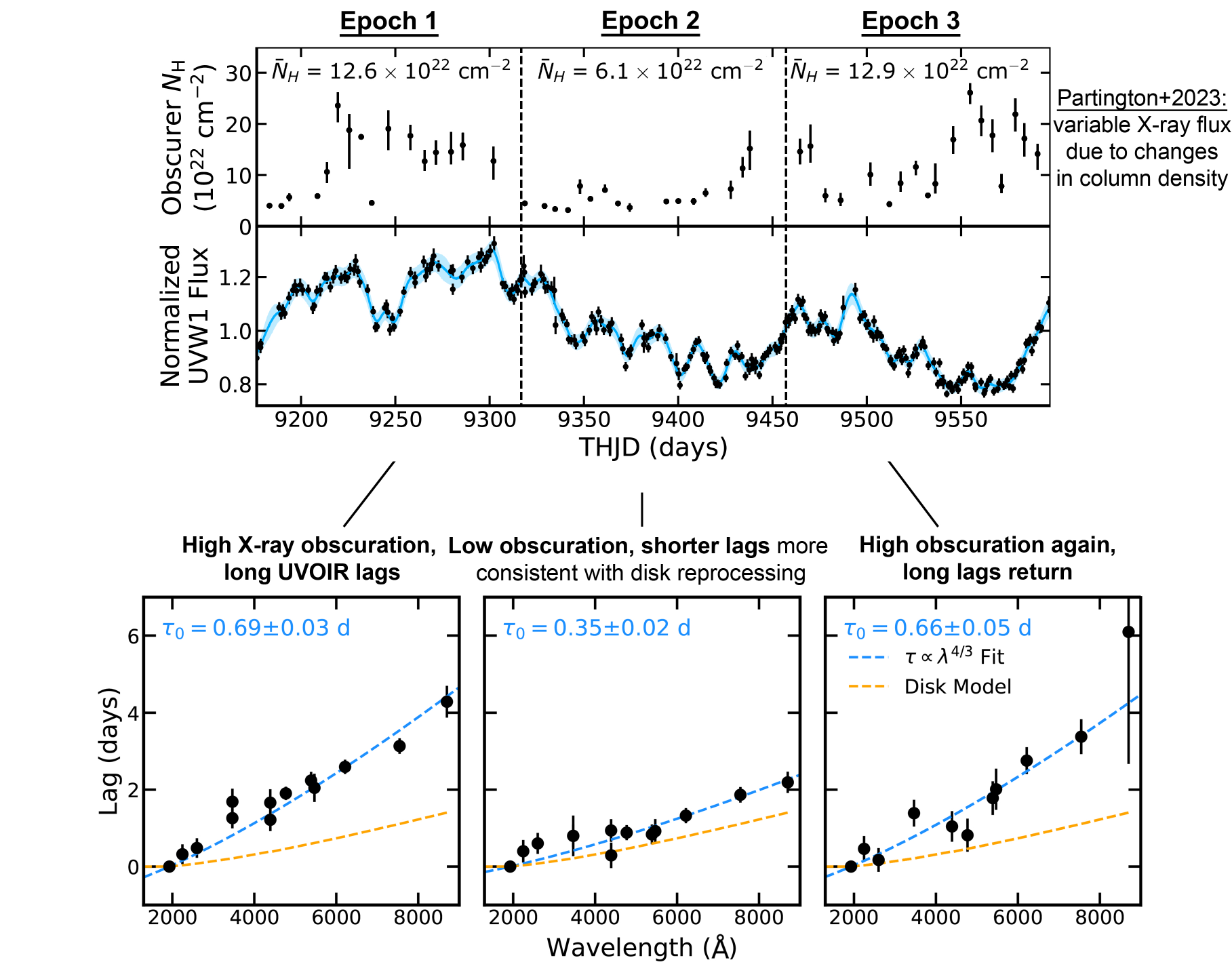}
    \caption{\textit{Top:} Obscurer column density over the course of the campaign from the NICER spectral analysis in Paper~III \citep{Partington_2023}. Dashed lines demarcate the three 140-day epochs in which we compute the lags. Epochs 1 and 3 coincide with times of higher average column density,  and Epoch~2 with times of lower average column density. \textit{Middle}: UVW1 light curve from Figure~\ref{fig:ibrm_lcs}. {Bottom}: Lags as a function of wavelength at frequencies of $0.014-0.048~\text{day}^{-1}$, equivalent to timescales of $20-70~\text{days}$. The lags are fit with a $\tau \propto \lambda^{4/3}$ relation (dashed blue line), with best-fit normalization values ($\tau_0$) shown. We compare the lags to the expected lag-wavelength relation given the mass and accretion rate of Mrk~817 using the disk reprocessing model in each frequency range (dashed orange line). The lags in Epochs~1 and 3 are a factor of 2 longer than those measured in Epoch~2.}
    \label{fig:lws_epoch}
\end{figure*}

\begin{figure*}[t!]
    \centering
    \includegraphics[width=\textwidth]{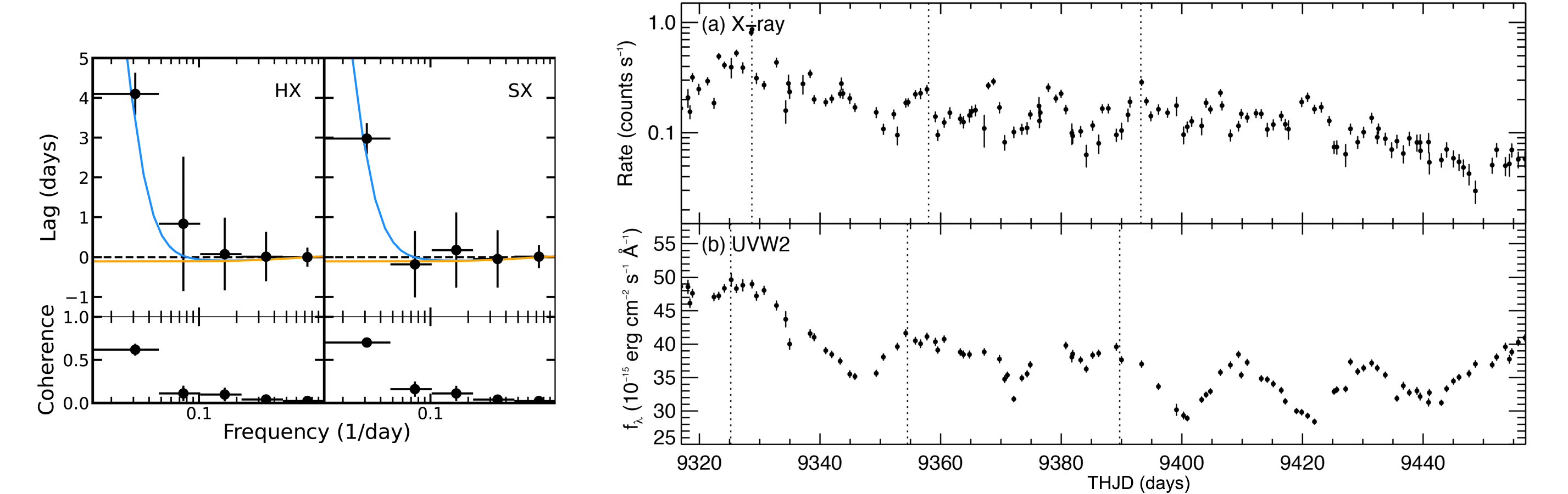}
    \caption{\textit{Left:} Lags and bias-subtracted coherence values as a function of frequency computed between the Swift X-ray bands (SX: 0.3--1.5~keV, HX: 1.5--10~keV) and the \textit{UVW2} band. A positive lag indicates that the X-ray variability lags the UV, opposite to the negative lag predicted by the disk model (orange). These lags can be reproduced by including a model component corresponding to reprocessing at a radius consistent with the BLR (blue). \textit{Right:} a) Swift 0.3--10~keV and b) \textit{UVW2} light curves. Dotted lines visualize the 3.5-day lag (the average of the SX and HX vs. \textit{UVW2} lags) between these two bands. Both bands exhibit a similar decrease in flux over time, which may be producing these measured lags on long timescales.}
    \label{fig:lfs_xray}
\end{figure*}
To study how the UVOIR lags evolve over the course of the 420-day Swift and ground-based campaign, we evenly split the campaign into thirds and computed the lags in the three resulting 140-day epochs delineated at THJD = 9317, 9457. For convenience, the epochs are labeled in the UVW1 light curve in the top panel of Figure~\ref{fig:lws_epoch}. This number of equal segments was selected to probe the same low-frequency range, given that the lowest frequency probed is set by the length of the observation. We also noticed that the epochs correspond approximately to periods of high and low column density of the obscurer reported from the NICER spectral analysis in Paper~III \citep{Partington_2023}. As shown in the top panel of Figure~\ref{fig:lws_epoch}, Epochs~1 and 3 coincide with periods of similarly high average column density ($\bar{N}_{\text{H}}/10^{22}~\text{cm}^{-2} = 12.6$ and $ 12.9$, respectively), whereas Epoch~2 instead overlaps with a period of lower average column density (lower by over a factor of 2 than the other epochs; $\bar{N}_{\text{H}}/10^{22}~\text{cm}^{-2} = 6.1$). The data quality does not vary dramatically across the three epochs, especially between Epochs~1 and 2, although notable gaps are present in the \textit{u, $B_g$, z} bands in Epoch~3. Given these large \textit{u, $B_g$} data gaps, we instead rely on the Swift \textit{U, $B$} bands in this epoch. We nonetheless demonstrate that simulated lags are successfully recovered in all epochs and wavebands (see Appendix~\ref{sec:sims}).

The low-frequency lags (at frequencies $0.014-0.048~\text{day}^{-1}$, equivalent to timescales of $20-70~\text{days}$\footnote{The low-frequency range of the epoch-resolved lags ($0.014-0.048~\text{day}^{-1}$) is very similar to one of the frequency bins of the full campaign lags ($0.02-0.05~\text{day}^{-1}$). Adjusting the epoch-resolved frequency range to be identical to the full-campaign bin produces results consistent within error.}) computed for each epoch are presented in the bottom panel of Figure~\ref{fig:lws_epoch}. These lags are also provided in the Appendix (Table~\ref{tab: epochresolved_lags}). The epoch-resolved lags show an expected anti-correlation between amplitude and frequency, but with notably larger uncertainties than the full-campaign lags, with data gaps taking up a larger fraction of the data.

We perform the same procedure as the lags computed using the full campaign: the low-frequency lags in each epoch are fit with a $\tau \propto \lambda^{4/3}$ relation (i.e., fitting $\tau = \tau_0 [(\lambda/\lambda_0)^{4/3} - 1]$ for the normalization $\tau_0$), as expected from standard accretion disk reprocessing \citep{Cackett_2007}. We also compare the lags to those predicted by the disk reprocessing model, as detailed in Section~\ref{sec:modeling}. 

The lags in higher-column-density Epochs~1 and 3 (left and right panels in Figure~\ref{fig:lws_epoch}) are generally consistent, producing best-fit normalizations of $\tau_0 = 0.69\pm0.03$ and $0.66\pm0.05~\text{days}$, respectively. These lags are longer than those predicted by the disk reprocessing model by a factor of 2--3. Similar to the lags computed from the full campaign, we also observe a \textit{U}-band lag excess in both the first and third epochs, where the lag surpasses the $\tau \propto \lambda^{4/3}$ best fit by roughly 110\% in Epoch~1 and 80\% in Epoch~3.

The lags in lower-column-density Epoch~2 (middle panel in Figure~\ref{fig:lws_epoch}) are systematically shorter than those measured in the other two epochs: the best-fit normalization of $\tau_0 = 0.35\pm0.02~\text{days}$ is smaller by over a factor of two than the normalizations of the other epochs, and $>10\sigma$ from that measured in Epoch~1. Again, the data quality in Epochs~1 and 2 is consistent in general, and we do not find systematically poorer recovery of simulated lags in this epoch. Unlike Epochs~1 and 3, however, the lags in Epoch~2 are nearly consistent with those expected from disk reprocessing. This includes a lack of the \textit{U}-band excess observed in the other epochs (the $\tau \propto \lambda^{4/3}$ fit lies within 1$\sigma$ from the \textit{U}-band lag in Epoch~2). We find in Section~\ref{sec:discuss} that if we extend the time range of Epoch~2 to include the first half of Epoch~3 where the column density is still low, the lags become fully consistent with the disk reprocessing model. The measured bias-subtracted coherences are high at these low frequencies, with an average coherence of 0.93 in Epoch~2 and 0.79 in Epochs~1 and 3. 

In order to evaluate how the variability changes across epochs, we computed the PSD in each UVOIR band per epoch using the GP realizations. The PSDs are then fit with a power law $P(\nu) = A\nu^\beta$ at frequencies $<0.1~\text{day}^{-1}$. These frequencies are low enough to avoid white noise while including timescales where we expect to see reprocessing by the disk. The best-fit PSD slopes ($\beta$) range from -3.5 to -2.4 depending on the band and epoch, which are similar to those found for other AGN \citep[][]{Edelson_2014, Panagiotou_2022a}. The average PSD slope in each epoch is $-3.3\pm0.2$ (epoch~1), $-2.7\pm0.3$ (epoch~2), $-2.9\pm0.4$ (epoch~3). This indicates  there is more power at high frequencies relative to low frequencies that in Epochs~2 and 3 than in the first epoch. It may be that the first half of Epoch~3, which we found contributes to shorter lags similar to those in Epoch~2 (see Figure~\ref{fig:lws_longer}), is contributing to a shallower slope of the epoch's PSD. Splitting Epoch~3 to test this, however, renders us unable to meaningfully constrain the slope of the Epoch 3 PSD. An advantage of a frequency-resolved analysis is that the underlying slope of the PSD does not systematically skew the measured lags and thus the inferred size scale; as is the case for the CCF approach, due to the variability amplitude being strongest on the longest timescales.

These significant changes in the lags from epoch to epoch provide a unique opportunity, especially to explore the origin of the perplexingly long lags reported from previous continuum reverberation mapping campaigns. In summary, we measure longer UVOIR lags at times of higher column density, similar to the UV emission line lags presented in Paper~V \citep{Homayouni_2023_2}. Potential explanations are discussed in Section~\ref{sec:discuss}. 

We measure a (Swift) X-ray-UV lag in only the second epoch, which is unsurprising given the low X-ray flux throughout the rest of the campaign. Figure~\ref{fig:lfs_xray} presents the X-ray-UV lags in both the soft (0.3--1.5~keV) and hard (1.5--10~keV) bands. Similar to the frequency-resolved lags in Mrk~335 \citep{Lewin_2023}, the X-ray \textit{lags} the UV at low frequencies; in this case, the soft and hard bands lag the UV by 3 and 4 days, respectively, in the $0.014-0.048~\text{day}^{-1}$ frequency range. This lag is visualized with the X-ray and \textit{UVW2 }light curves shown in the same figure. While there are some notable features that could explain the lag, it is also possible that the lags are being produced by the overall (long-timescale) negative slope present in both bands.

These lags appear contrary to the central reprocessing picture, in which the coronal X-rays are reprocessed by circumnuclear material to generate delayed variability at longer wavelengths. In Mrk~335, the positive X-ray lag was explained by reprocessing by photoionized gas near the BLR that dominates the X-ray spectrum below 2~keV \citep{Lewin_2023}. We are similarly able to reproduce the observed X-ray-UV lags with additional reprocessing from beyond the disk at the same radius consistent with the BLR used to describe the UVOIR lags. Nonetheless, the lags may be the result of propagating fluctuations in the mass-accretion rate \citep{Arevalo_2006} or reprocessing by the obscuring disk wind positioned at $\sim$4 light-days from the black hole \citep{Zaidouni_2024}.

\section{Modeling the UVOIR Lags} \label{sec:modeling}

We aim to model the frequency-resolved UVOIR lags of Mrk~817 presented in the previous section. We are particularly interested in 1) the agreement (or lack thereof) between the measured lags and disk reprocessing and 2) the geometry and degree of additional reprocessing required to reproduce the measured lags. We model the anticipated disk reprocessing lags by applying the model of \cite{Cackett_2007}. The model is parameterized by the temperatures of the disk at an arbitrary radius of 1 light-day during a faint state and a bright state ($T_B, T_F$).  The CCF lags depend on these temperatures \citep[Equation~13 in][]{Cackett_2007}; as such, we determine the temperatures by fitting them to the (CCF) lag-wavelength relation expected from standard disk reprocessing \citep[Equation~12 in][]{Fausnaugh_2016}. Assuming a black hole mass of $M_{\text{BH}} = 3.85\times 10^7~\text{M}_\odot$ and an Eddington ratio of $L_{\text{Bol}}/L_{\text{Edd}} = 0.2$ for Mrk~817, a lag normalization of $\tau_0 = 0.31~\text{days}$ is anticipated. Reproducing this normalization with the \cite{Cackett_2007} model results in temperatures of $T_F=5900~\text{K}$ and $T_B=11400~\text{K}$.

In order to compute the frequency-resolved lags, we first generate the impulse-response function for each waveband using Equation~7 in \cite{Cackett_2007}. We assume an inclination of $19^\circ$ \citep{Miller_2021}, although changes in inclination have a minor effect the model. The transfer function\footnote{As a reminder, the transfer function is obtained by taking the Fourier transform of the impulse-response function.} in each band of interest is then multiplied by the complex conjugate of the reference band (\textit{UVW2}) transfer function to account for the reference band also being a reprocessed light curve \citep[for details, see][]{Cackett_2022}. This product of transfer functions is a cross-spectrum, and thus the phase lag is converted to time lag by dividing by $2\pi\nu$, where $\nu$ is the frequency of the bin.

The resulting lags from the disk reprocessing model are shown as a function of frequency in Figure~\ref{fig:lfs_full}. Moving from high-to-low frequencies is equivalent to moving outwards in the disk: the model predicts longer lags at larger radii until reaching a radius of maximum reprocessing, beyond which reprocessing becomes negligible and the lag becomes constant (at low frequencies). 

\subsection{Modeling the full campaign lags}

\begin{figure}[t!]
    \centering
    \includegraphics[width=\columnwidth]{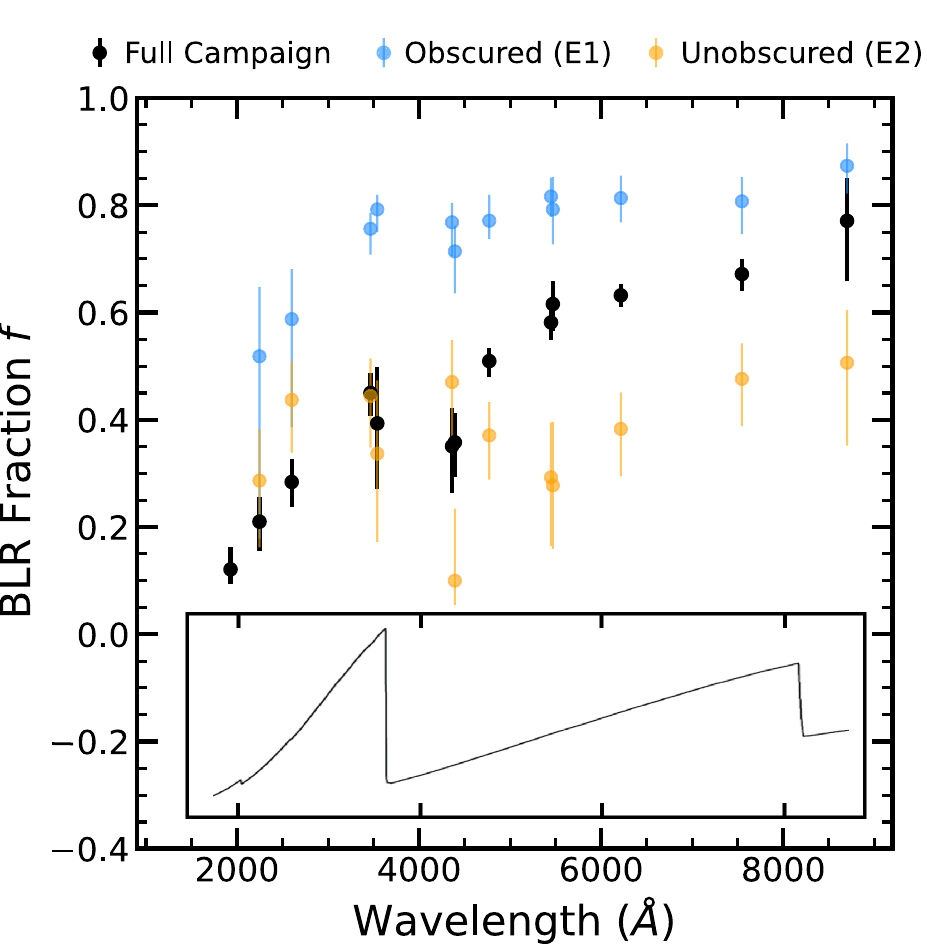}
    \caption{Fraction of the total impulse-response function model made up the BLR component. The BLR contributes more to the total model during times of high obscuration (Epoch~1, blue) than when obscuration is low (Epoch~2, orange), with the full-campaign BLR fractions most often between those of these two epochs. The BLR fractions for Epoch 1 and the full campaign show a local maximum in the \textit{U} band near 3500 \AA, consistent with the Balmer jump in the diffuse line and continuum. For comparison, the inlay is Figure~9 from \cite{Korista_2019}, which displays the ratio of the diffuse line and continuum emission to the total SED as a function of wavelength, with x-axis aligned to match our plot.}
    \label{fig:blrfrac}
\end{figure}
The disk model is consistent with the lags at high frequencies ($>0.05~\text{day}^{-1}$), but undershoots the lags at low frequencies by a factor of 3--4 on average. The reduced chi-squared of the disk model is $\chi^2_\nu = 375.6/72 = 5.2$, which does not include the X-ray bands (we  observe a significant lag only in the epoch-resolved lags). The model is qualitatively different from the observed lags at frequencies below $0.05~\text{day}^{-1}$: the disk model predicts a constant lag, whereas the lags continue to increase at lower frequencies. This indicates additional reprocessing at larger radii not being accounted for by the disk model. 

These results are similar to those found from modeling the frequency-resolved lags in NGC~5548 \citep{Cackett_2022} and Mrk~335 \citep{Lewin_2023}: the low-frequency lags cannot be reproduced with disk models alone, even with higher temperatures producing larger effective radii of reprocessing. Instead, the lags are successfully modeled by including an additional impulse-response function, corresponding to reprocessing from a more-distant reprocessor. In both sources, the best-fit radius of this distant reprocessor is consistent with the BLR size scale inferred from previous H$\beta$ measurements. 

We apply this procedure in an attempt to model our lags using a phenomenological log-normal impulse response function as in \cite{Cackett_2022} and \cite{Lewin_2023}:
\begin{equation}
    \psi_{\text{BLR}}(t) = \frac{1}{S\sqrt{2\pi}t}\exp \left[ -\frac{(\ln(t)-M)^2}{2S^2} \right]
\end{equation}

\noindent where the median ($e^M$) and standard deviation ($S$) are determined by minimizing the total chi-squared. While this is a phenomenological model, it can be interpreted as reprocessing from beyond the disk, and gauges the geometry (radius and width) of the distant reprocessor required to reproduce the observed lags. While the standard deviation of the component is a proxy for the width of the reprocessor, the two are not interchangeable. For context, this log-normal impulse response is a smoother alternative to the popular top-hat response function used to model reprocessing by a spherical shell \citep{Uttley_2014}. The log-normal response function is also asymmetric, with a tail to long lags.

The final impulse-response function is a linear combination of the disk model ($\psi_{\text{disk}}$) and the distant reprocessor model ($\psi_{\text{BLR}}$):
\begin{equation} \label{eq:totresp}
    \psi_{\text{tot}}(t) = (1-f)\psi_{\text{disk}}(t)+f\psi_{\text{BLR}}(t)
\end{equation}

\noindent where $f$ is the fractional contribution of the distant reprocessor component to the final model. We fit for this fraction in each band independently, akin to the BLR diffuse line and continuum varying across wavebands \citep{Korista_2001, Korista_2019, Lawther_2018, Netzer_2022}. For clarity, the BLR fraction does not represent the fraction of the observed flux originating from the BLR. Instead, the model parameter has a non-linear relationship with the amount that the observed lags exceed the disk model. Modeling the lags in NGC~5548 and Mrk~335 in this way produced $f$ values as a function of wavelength that are qualitatively similar to the BLR continuum \citep[Figure~7 and 8 in][respectively]{Cackett_2022, Lewin_2023}. This includes higher $f$ values (i.e., the distant reprocessor contributed more to the total model) required in the \textit{U} band versus neighboring bands.

The final disk+BLR model (shown in Figure~ \ref{fig:lfs_full}) is a significant improvement in reproducing the lags at low frequencies compared to the disk model: the reduced chi-squared is improved to $\chi^2_\nu = 47.9/57 = 0.84$ from $\chi^2_\nu = 5.2$ in the case of the disk model. The median of the BLR model---a proxy for the size scale of additional reprocessing needed to reproduce the lags---is constrained to $e^M = 22.5\pm3.6~\text{days}$. This radius is consistent with that of the BLR based on H$\beta$ lag measurement from Paper~I \citep[$\tau_{\text{H}\beta}=23.2\pm1.6~\text{days}$;][]{Kara_2021}. The standard deviation of the component is constrained to $S=4.3\pm0.8~\text{days}$. This value is larger by over a factor of 4 than that found when the same modeling treatment was applied to the frequency-resolved lags in NGC~5548 and Mrk~335 \citep{Cackett_2022, Lewin_2023}. In Section~\ref{sec:discuss}, we demonstrate that this larger standard deviation can be produced as a result of averaging over lags produced at different radii in different epochs.

The best-fit BLR fractions are presented in Figure~\ref{fig:blrfrac}. Similar to NGC~5548 and Mrk~335, the shape of the BLR fraction as a function of wavelength shows several similarities to that of the BLR diffuse line and continuum (overlaid in the bottom of the same Figure~for reference). Most notable is a local maximum in the \textit{U} band, a change in slope in before and after the Balmer jump, and larger BLR fractions at longer wavelengths \citep{Korista_2001, Korista_2019}.

\subsection{Modeling the epoch-resolved lags}
The UVOIR lags show systematic changes between the three epochs defined by splitting the 420-day light curves evenly into three segments. Epochs 1 and 3 exhibit similar (low-frequency) lags (see Figure~\ref{fig:lws_epoch}), producing lag normalizations within $1\sigma$ of each other. The lags in Epoch~2, however, are shorter than those in the other epochs by over a factor of 2 on average. 

We apply the same procedure as the full campaign lags to the lags in each epoch: we evaluate the performance of the disk model, and then include the distant reprocessor component to constrain the geometry (namely, the median and standard deviation in radius) of the reprocessor needed to describe the low-frequency lags. As shown in Figure~\ref{fig:lws_epoch}, the shorter lags in Epoch~2 are nearly consistent with the disk model, whereas the lags in Epochs~1 and 3 exceed the disk model by a factor of $\sim$3 on average. This is reflected by the reduced chi-squared values totaled across all UVOIR bands and frequencies: $\chi^2_\nu = 48.3/60 = 0.8$ in Epoch~2, versus $\chi^2_\nu = 416.2/60 = 6.9$ and $\chi^2_\nu = 104.8/50 = 2.1$ in Epochs~1 and 3, respectively (fewer degrees of freedom as the \textit{u, $B_g$} bands are not used in Epoch~3). 

We then include the log-impulse distant reprocessor model component, fitting for the median and standard deviation in each epoch independently. The reduced chi-squared significantly improves in Epochs~1 and 3: $\chi^2_\nu = 35.1/45 = 0.78$ (Epoch~1) and $\chi^2_\nu = 26.1/37 = 0.70$ (Epoch~3). The fit statistic for Epoch~2 improves slightly to $\chi^2_\nu = 32.0/45 = 0.71$, as expected given the low contribution of the BLR component needed to describe the lags. The best-fit model parameters agree within $1\sigma$ across epochs, suggesting that the same distant reprocessor is contributing to the low-frequency lags in all epochs. The best-fit medians are all consistent with the BLR size scale inferred from the 23-day H$\beta$ lag \citep{Kara_2021}, with $e^M = 22.3\pm6.3~\text{days}$ in Epoch~1, $e^M = 24.3\pm5.8~\text{days}$ in Epoch~2, and $e^M = 24.0\pm7.4~\text{days}$ in Epoch~3. The best-fit standard deviations of the distant reprocessor are $S = 0.9\pm0.7$ in Epoch~1, $S = 1.7\pm0.9$ in Epoch~2, and $ S = 1.9\pm1.2$ in Epoch~3. These values are over a factor of 2 smaller than that required to describe the full-campaign lags ($S=4.3\pm0.8~\text{days}$), and are similar to previous values to model the lags in NGC~5548 \citep[$S=1.1\pm0.2$;][]{Cackett_2022} and Mrk~335 \citep[$S=0.9^{+0.2}_{-0.1}~\text{days}$;][]{Lewin_2023}. These statistics are used in the next section to demonstrate that the larger standard deviation of the BLR component required to describe the full-campaign lags can be reproduced by averaging over lags produced at different effective radii in different epochs. 

The similar lags in Epochs~1 and 3 require the BLR model to compose a high fraction of the total model, with an average BLR fraction of $f = 0.74\pm0.09, 0.70\pm0.14$ in Epochs~1 and 3, respectively. The BLR fraction in Epoch~2 is lower by almost a factor of 2 ($f = 0.40\pm0.11$), as expected given that the low-frequency lags in this epoch are nearly consistent with disk reprocessing. In Section~\ref{sec:discuss}, we discuss that line-of-sight obscuration by a disk wind may be obscuring our view of reprocessing by the disk more significantly in Epochs~1 and 3, producing longer lags from instead viewing mostly reprocessing at larger radii, including the BLR. In this scenario, we expect higher values for the BLR fraction in Epochs~1 and 3 given that the median response radius of the reprocessor is not changing significantly, which is consistent with the results.

The hard and soft (Swift) X-ray bands were found to lag the \textit{UVW2} band by $4.1\pm0.5$ and $3.0\pm0.4$ days, respectively, in the lowest frequency bin in Epoch~2 (see Figure~\ref{fig:lfs_xray}). This indicates that the hard X-rays lag the soft by $\sim$1 day. A simple order-of-magnitude comparison to the XMM-Newton hard lag reveals that the lags are consistent after adjusting for the difference in frequency: $\tau = 10^5~\text{s}, \nu= 10^{-8}~\text{Hz}$ for Swift versus $\tau = 10^2~\text{s}, \nu= 10^{-5}~\text{Hz}$ for XMM-Newton. This assumption is motivated by theoretical work for the frequency-amplitude scaling of inward propagating fluctuations in the mass-accretion rate \citep{Lyubarskii_1997, Kotov_2001, Arevalo_2006}, and suggests that the lags may originate from this same physical process.

We also modeled the X-ray-UV lags on their own to gauge the region of reprocessing capable of reproducing these positive lags. The best-fit parameters of the BLR component are consistent with those constrained with the UVOIR lags: the median/effective radius is $e^M = 25.0\pm8.7~\text{days}$, with a standard deviation of $S = 2.5\pm1.5~\text{days}$. Nonetheless, the coherence is moderate ($\sim$0.5-0.7), and more notably that the signal-to-noise ratio is low in these bands. As such, these results are tentative, albeit similar to Mrk~335, where the soft X-rays were also measured to lag the UV \citep{Lewin_2023}. It is nonetheless worth mentioning that the variability we see in this low-column-density epoch more directly reflect variations in the intrinsic X-ray continuum rather than in the transparency of the obscuration.

\section{Discussion} \label{sec:discuss}

\subsection{Contamination of the disk continuum reverberation lags by the broad-line region}
The lag measurements from recent, high-cadence monitoring campaigns of AGN using Swift and ground-based telescopes have consistently deviated from theoretical predictions. First, the normalizations of the expected $\tau\propto\lambda^{4/3}$ relation are typically a factor of 2--3 larger than predicted given the black hole mass and accretion rate. The lags have been especially long in the \textit{U} band (3465~\AA), exceeding the best-fit $\tau \propto \lambda^{4/3}$ relation even with the large normalizations by a factor of $\sim$2 \citep{Cackett_2018, Edelson_2019, HernandezSantisteban_2020, Vincentelli_2021, Kara_2022}. These \textit{U}-band excesses are commonly attributed to the diffuse line and continuum of the BLR \citep{Korista_2001, Korista_2019, Lawther_2018, Netzer_2022}. 

The lags at frequencies below $0.05~\text{day}^{-1}$ are longer than those expected from disk reprocessing given the mass and accretion rate of Mrk~817. The lags at frequencies above $0.05~\text{day}^{-1}$ are instead well described by the disk reprocessing model, including the lag in the \textit{U}-band. If the discrepancies from disk reprocessing at low frequencies are due to contamination from a distant reprocessor, then the reprocessing is occurring on timescales longer than $1/0.05 = 20~\text{days}$. This timescale is consistent with the outer parts of the BLR according to the 23-day H$\beta$ lag in Paper~I \citep{Kara_2021}. Nonetheless, a small lag excess remains above this frequency beyond 6000~\AA\ that resembles the Paschen continuum, suggesting that BLR reprocessing continues to smaller radii. This is consistent with the extent of the BLR inferred from the lags in Paper I, which range from the UV to H$\beta$ and span from a few light days to beyond 23 light days.

If we instead fit the disk model temperatures to the measured lags in the lowest frequency bin, an Eddington ratio an order of magnitude higher than observed is required \citep[$L/L_{\text{Edd}}=6.7$ vs. $0.2$;][]{Kara_2021}. The assumptions of the thin-disk model are not expected to hold at this accretion rate \citep{Netzer_Trak_2014, Wang_2014, Du_2018}. To avoid this unlikely conclusion, we included a simple log-normal impulse-response function to account for additional reprocessing from beyond the disk. The inclusion of this component significantly improves the description of the lags ($\chi^2_\nu = 0.84$ vs. $\chi^2_\nu = 5.2$ from the disk model alone). 

Similar to the modeling results in Mrk~335 and NGC~5548, the required median radius of the component ($22.5\pm3.6~\text{days}$) is consistent with the 23-day H$\beta$ lag \citep{Kara_2021}. Photoionization models have shown that the mean emissivity radius of the diffuse continuum is smaller than that of H$\beta$ by a factor of $\sim$2--3 \citep{Korista_2019, Netzer_2020, Netzer_2022}. Given that the log-normal distribution is right-skewed, the distribution peaks at a value smaller than the median; in other words, the radius of maximum reprocessing is smaller than the median value. A median of 23~days and a standard deviation of 1~day found from fitting the individual epochs places the radius of maximum reprocessing at 8.2 days, with significant reprocessing down to a few days. This radii range is consistent with the lags from Paper I indicative of the size scale of the BLR and the aforementioned modeling by \cite{Korista_2019}.

The standard deviation of this additional component, which is related to the width of the extended reprocessor, is constrained to $4.3\pm0.8~\text{days}$. This value is over a factor of 4 larger than that required to describe the lags in NGC~5548 and Mrk~335 \citep[using the same log-normal and disk model;][]{Cackett_2022, Lewin_2023}. Modeling the lags in each epoch results in a smaller standard deviation ($\sim1~\text{day}$). This suggests that the larger standard deviation may be the result of combining epochs across which the lags differ notably, for instance due to reprocessing at different radii in different epochs. This idea agrees with the full-campaign lags being roughly an average of the shorter lags in Epoch~2 and longer lags in Epochs~1 and 3. To test this idea, we compute an effective radius of reprocessing for each epoch (in each waveband) by using the best-fit BLR fraction ($f$) and radius ($R_{\text{BLR}}$): $R_{\text{eff}} = (1-f)R_{\text{disk}} + fR_{\text{BLR}}$. We then compute the standard deviation of the effective radii across the three epochs to compare to the large standard deviation required to model the full campaign lags. The radius of disk reprocessing ($R_{\text{disk}}$) depends on wavelength and is computed using Equation~12 in \cite{Fausnaugh_2016}. We set $R_{\text{BLR}} = 23~\text{days}$ to match the H$\beta$ lag, which is consistent with the best-fit median in all epochs. The resulting standard deviation of effective radii across the three epochs ranges from 2.5--6~days, depending on the band, or $4.1\pm1.0~\text{days}$ on average. This is consistent with the larger standard deviation needed to reproduce the full-campaign lags, thus supporting this value could be the result of averaging over lags produced at different effective radii in different epochs. Specifically, the effective radii of reprocessing averaged across bands are 17.5~days (Epoch~1), 9.1 days (Epoch~2), and 13.0 days (Epoch~3).

We caution that our modeling of the lags is limited: we apply a single model for disk reprocessing and a simple analytic treatment for the extended reprocessor/BLR. As such, additional modeling with more physical models is warranted, such as those that include a more-realistic treatment of the coronal geometry \citep{Kammoun_2021a, Kammoun_2021b}, the thickness of disk \citep{Starkey_2023}, and models for the BLR \citep{Korista_2019, Netzer_2020}. Modeling of the lags from inwardly-propagating fluctuations in the mass-accretion rate \citep[e.g.][]{Lyubarskii_1997, Neustadt_2024} is also of importance.

\begin{figure*}[t!]
    \centering
    \includegraphics[width=\textwidth]{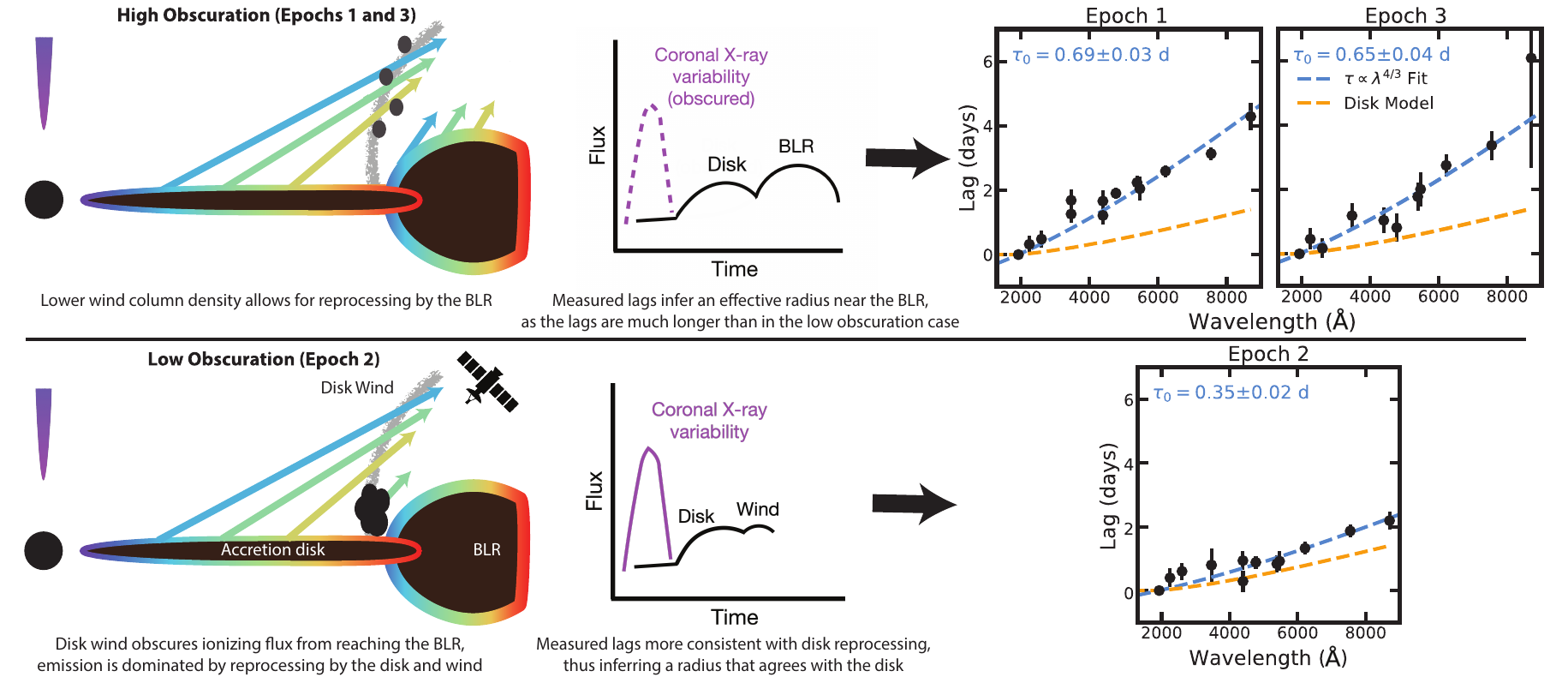}
    \caption{One possible explanation for the longer UVOIR lags at times of higher column density involves a disk wind, as informed by spectral analysis in Papers III, IX \citep{Partington_2023, Zaidouni_2024}. At times of high column density \textit{(upper)}, the wind is launched into our line of sight. We measure long continuum lags because the BLR is exposed to the ionizing source, contributing to longer lags. These additional regions of reprocessing respond on longer timescales than the disk and produce a \textit{U}-band excess (\textit{Middle}: demonstrative light curves, with obscured light indicated by dashed lines). At times of low column density \textit{(lower)}, the dense base of the wind intercepts the ionizing flux from reaching the BLR. The measured lags are dominated by disk and wind reprocessing, resulting in shorter lags that lack a significant \textit{U}-band excess.}
    \label{fig:epoch_schem}
\end{figure*}
\subsection{The connection between time lags and the obscurer}

There are several key observables presented here and in previous AGN STORM 2 papers that need to be explained:
\begin{itemize}
    \item Paper~III \citep{Partington_2023} found the line-of-sight X-ray column density and the equivalent width of the blueshifted broad UV absorption troughs vary together on timescales of $\sim$100 days. Here, we separated the observations into three epochs corresponding to these periods of high (Epochs~1 and 3) and low (Epoch~2) column density.
    \item Paper~IX \citep{Zaidouni_2024} found concurrent X-ray and UV absorption lines with the same velocity profile, i.e. from the same outflow. It is estimated that the launching radius of the outflow is roughly 1000 $R_g$ (2-3 light days) and estimated that the time it takes the wind to travel from the disk into our line of sight is 100-200 days.
    \item Paper~IV \citep{Cackett_2023} found that while the \textit{UVW2} and HST emission at shorter wavelengths is usually highly correlated, there is a relative excess of the \textit{UVW2} emission compared to all HST bands when the column density is high during the campaign corresponding to our Epoch~1.
    \item Paper~V \citep{Homayouni_2023_2} found that the \ion{C}{4} lag changes dramatically over time, with the longest lag occurring when the UVW2/HST correlation weakens and the column density is relatively high in Epoch~1. 
    \item Here, we find the continuum lags vary significantly from epoch to epoch. The lags are over a factor of 2 larger when the column density is high (Epochs~1 and 3) than when the column density is low (Epoch~2).
\end{itemize}

The results presented in this paper demonstrate that the continuum lags can be strongly affected by the presence of a variable obscurer. While a holistic analysis that includes modeling changes in the emission and obscuration is required to fully characterize the lags, here we present one possible explanation, illustrated in Figure~\ref{fig:epoch_schem}. In this physical picture, the epoch-to-epoch changes in the lags are due to changing obscuration by an accretion disk wind, launched at $\sim 1000~R_g$. We can think of this as a limit cycle, where the wind is first launched from the disk, and has a high column density. The wind is lifted from the disk, becoming more clumpy and lower column density, as it reaches our line of sight. This wind moving from the disk surface and up into our line of sight explains the changes in X-ray column density and UV absorption troughs on timescales of 100--200 days (Paper III \& Paper IX).

In the context of our delineated `epochs’, the disk limit cycle begins with the high column density wind being launched from the disk in Epoch~2. This wind intercepts the radiation to the BLR, thus leading to relatively short lags\footnote{This interpretation that the observed continuum lags become shorter with decreasing column density is further supported if we extend Epoch~2 to include the first part of Epoch~3, when the column density was still relatively low (considering THJD = 9317--9525). As shown in Figure~\ref{fig:lws_longer}, the lags become even shorter than those from the original Epoch~2 and are fully consistent with the disk reprocessing model, including a resolution of the U-band excess.}. As the wind is lifted from the disk’s surface and into our line of sight (Epochs~1 \& 3), it becomes more clumpy and lower column density, and the BLR is exposed to the ionizing source, thus producing longer lags that are indeed dominated by the diffuse BLR continuum. It is also in these Epochs where the longest \ion{C}{4} lags are observed (Paper~V).

This model is appealing in that it fits within the larger context of UVOIR campaigns, where in all the unobscured AGN, we observe lags that are ubiquitously longer than predicted by disk reprocessing by a factor of 2 or more. It is for this reason that studies have invoked more distant reprocessing from the diffuse BLR (e.g. \citealt{Korista_2001, Lawther_2018, Korista_2019, Netzer_2022}). However, here in Epoch~2 in the campaign of Mrk~817, we observe (to our knowledge, for the first time), lags that are fully consistent with expectations from disk reprocessing alone. Perhaps it is precisely because this source has a time-variable obscurer that at some unobscured epochs we see short lags more consistent with just the disk.

There are important details to modeling that will be discussed in Netzer et al., in prep. For instance, because the wind base is a high column density, photoionized plasma, it is expected to radiate its own reprocessed emission at a radius of a few light days from the black hole. In this model, in Epoch~2, the lags are dominated by the re-radiation from the wind base, that is still effectively located at a smaller radius than the diffuse BLR.

While this model explains many of the observations from Papers~I-IX, there are some caveats that require future work. For instance, it remains unclear how the emission-line fluxes fit into the picture. Paper~V shows the \ion{C}{4} flux has its strongest response relative to the continuum when the \ion{C}{4} and continuum lags are shortest (Epoch~2, Phase~C in Paper~V). Yet this result implies a strong contribution to continuum lag measurements from the BLR, contrary to our observation of short continuum lags consistent with primarily disk emission. It is difficult to imagine how this picture could produce the short lags that are fully consistent with reprocessing by just the disk at times of low column density, given that BLR reprocessing is present at all times in the Paper~V scenario. Moreover, one might expect that in Epoch 2, where, in our model, the obscurer does not affect the continuum as much, and the lags do not require contributions from the BLR, we would expect a larger correlation between X-rays and UVOIR, but this is not observed. Some of this is likely because the obscuration is still present (albeit at a lower level), but may also require examination of effects of dynamic variability of the X-ray corona \citep{Panagiotou_2022b} and/or fluctuations in the disk properties in response to FUV or X-ray heating \citep{Gardner_2017, Sun_2020, Chen_2024, Secunda_2024}.

While this physical model does broadly explain many of the main results of the AGN STORM 2 campaign, a more holistic approach to modeling the changes in emission and obscuration are needed in order to disentangle contributions from the disk, obscurer, and BLR. The specifics of a more-complete model predicting the time lags together with the emission/absorption line kinematics will be examined in detail in forthcoming works. 

\section{Conclusions} \label{sec:conclusions}
We present the frequency-resolved lags of AGN Mrk~817 computed from 14~months of high-cadence data by Swift and ground-based telescopes, in addition to an XMM-Newton observation, as part of the AGN STORM 2 campaign. For the first time, X-ray reverberation lags, which probe the innermost accretion disk, are detected during a simultaneous UVOIR disk reverberation mapping campaign, which reveal reprocessing all the way out to the outer disk and the BLR---thus, producing a (reverberation) map effectively spanning the entire accretion disk and the inner accretion flow. Here are the main results:

\begin{enumerate}
    \item The XMM-Newton lags (see Figure~\ref{fig:xray_reverb}) reveal the first detection of a soft lag in this source. Both the frequency and amplitude of this lag are consistent with reverberation by the innermost accretion flow based on lag-mass scaling relations \citep{DeMarco_2013}. The lags at lower frequencies instead exhibit canonical hard lags commonly attributed to propagating fluctuations in the accretion rate.
    \item The Swift/ground-based lags in the UVOIR bands computed using the full campaign light curves are longer than those expected from standard disk reprocessing at low frequencies, by over a factor of 3 on average ($<0.05~\text{day}^{-1}$, i.e., timescales longer than 20 days; see the two left-most panels in Figure~\ref{fig:lws_full}). The lag discrepancy in the \textit{U} band is especially large, exceeding the best-fit $\tau\propto\lambda^{4/3}$ relation by a factor of 2. These discrepancies are similar to those reported in previous continuum reverberation mapping campaigns.
    \item At higher frequencies ($>0.05~\text{day}^{-1}$, i.e., timescales shorter than 20 days), the UVOIR lags are instead well described by standard disk reprocessing, including a resolution of the \textit{U}-band excess. In other words, the lag discrepancies resolve at timescales shorter than the 23-day H$\beta$ lag indicative of the BLR size scale, supporting that reprocessing from the BLR at farther radii is ``contaminating" the disk reprocessing lags. 
    \item  The UVOIR lags are well described when an additional model component accounts for reprocessing from a distant reprocessor (see Figure~\ref{fig:lfs_full}). The best-fit radius of the reprocessor at $22.5\pm3.6~\text{days}$ is consistent with the $\sim$23-day H$\beta$ lag measured in Paper~I \citep{Kara_2021}. The required width of the reprocessor ($\sim4~\text{days}$) is substantially larger than that measured from previous lag modeling, which may be the result of averaging over lags produced at different effective radii in different epochs.
    \item The UVOIR lags computed when splitting the campaign into three 140-day epochs show substantial variations. The lags in Epochs~1 and 3 are generally consistent, but longer than those in Epoch~2 by over a factor of 2 (see Figure~\ref{fig:lws_epoch}). According to the NICER spectral analysis in Paper~III \citep{Partington_2023}, the average column density of the obscurer in Epochs~1 and 3 is higher than that in Epoch~2. We suggest that, when the obscurer is strongest, additional reprocessing by the BLR elongates the lags. As the wind weakens, the lags are instead dominated by shorter accretion disk lags as the re-emerging, dense base of the wind shields the BLR from the ionizing source.
\end{enumerate}

\facilities{XMM-Newton, Swift, LCOGT, Liverpool:2m, Wise Observatory, Zowada, CAO:2.2m, YAO:2.4m}

\begin{acknowledgments}
This work makes use of observations from the Las Cumbres Observatory global telescope network. C.L., E.K. and F.Z. acknowledge NASA grants 80NSSC22K1120 and 80NSSC22K0570. M.C.B. gratefully acknowledges support from the NSF through grant AST-2009230. J.G. gratefully acknowledges support from NASA under the award 80NSSC22K1492. Y.H. was supported as an Eberly Research Fellow by the Eberly College of Science at the Pennsylvania State University. D.I., A.B.K and L.\v{C}.P. acknowledge funding provided by the University of Belgrade - Faculty of Mathematics (contract No. 451-03-47/2023-01/200104) and Astronomical Observatory Belgrade (contract No. 451-03-47/2023-01/200002) through the grants by the Ministry of Science, Technological Development and Innovation of the Republic of Serbia. A.B.K. and L.\v{C}.P. thank the support by Chinese Academy of Sciences President’s International Fellowship Initiative (PIFI) for visiting scientist. H.L. acknowledges a Daphne Jackson Fellowship, sponsored by the Science and Technology Facilities Council (STFC), UK. Research at UC Irvine was supported by NSF grant AST-1907290. E.M.C. gratefully acknowledges support from NASA through grant 80NSSC22K0089.  E.M.C. and J.A.M. gratefully acknowledge support from the National Science Foundation through AST1909199. M.V. gratefully acknowledges financial support from the Independent Research Fund Denmark via grant number DFF 8021-00130.y
\end{acknowledgments}

\appendix
\setcounter{figure}{0}
\renewcommand{\thefigure}{A\arabic{figure}}

\section{Assuming a normal flux distribution} \label{sec:normal}

Here, we assess whether we can assume Gaussianity of the observed Swift and ground-based flux distributions in order to model the observed variability using GPs. In the case that the empirical fluxes instead follow a \textit{log-normal} distribution \citep[as reported by previous AGN timing analyses, e.g.,][]{Uttley_2005}, we would train the GP on the log-transformed flux values and then exponentiate the realizations drawn from the resulting conditional posterior. The Cram\'er-von Mises (C-vM) test \citep{CvMtest} evaluates the goodness of fit of the empirical cumulative distribution function (ECDF) by that of the null hypothesis. For context, this test is an alternative to the more popular Kolmogorov-Smirnov (K-S) test \citep{Massey_1951}, but is more sensitive to small-scale differences in the ECDF \citep{2006ASPC..351..127B}. For each Swift and ground-based light curve, we perform two C-vM tests: one using a normal distribution null hypothesis, and the other using a log-normal null hypothesis. 

For the X-ray bands, only the normal null hypothesis C-vM tests result in \textit{p}-values less than 0.01 (hard band: $p\sim10^{-3}$, soft band: $p\sim10^{-6}$), indicating a low probability that the data follows a normal distribution. The log-normal tests in these bands, however, do produce \textit{p}-values greater than 0.01, indicating that we cannot reject that the X-ray fluxes are log-normally distributed at the 1\% confidence level. For the UVOIR bands, the C-vM tests produce \textit{p}-values ranging from 0.01 to 0.36, with a mean p-value of $\bar{p}=0.10\pm0.07$ for the normal null hypothesis tests and $\bar{p}=0.16\pm0.10$ for the log-normal null hypothesis tests. While neither null hypothesis can be rejected at the 1\% confidence level, the log-normal tests result in higher \textit{p}-values on average. As a result, we choose to train the GPs on the log-transformed flux values in all bands.

\section{Selecting the kernel function form} \label{sec:kernels}

There are infinite possible kernel function forms, as kernel functions are closed under addition. This motivates thorough (potentially unlimited) investigation when modeling new variability/processes \citep[e.g., creating kernel functions using deep neural networks;][]{2015arXiv151102222W}. In our case, previous work \citep[e.g.,][]{Wilkins_2019, Griffiths_2021, Lewin_2022, Lewin_2023} has tested the efficacy of several kernel function forms for modeling AGN variability for a wide range of data quality, including for Swift campaigns with similar data quality to ours \citep{Griffiths_2021, Lewin_2023}. We evaluate the same three kernel function forms as the aforementioned works: the squared exponential (SE), rational quadratic (RQ), and Mat\'ern-$\frac{1}{2}$ (hereafter M-$\frac{1}{2}$) kernels. We refer to \cite{Wilkins_2019} for an introduction to these functional forms.

As introduced in Section~\ref{subsec:gps}, we optimize the kernel function hyperparameters by maximizing the likelihood of the observed data given the GP model (i.e., minimize the NLML). We compare the efficacy of these kernel function forms by comparing 1) the optimized NLML values and 2) the mean-squared error (MSE) when the model predicts the last 20\% of the light curve after training (i.e., optimizing the hyperparameters) using the first 80\% of the light curve. We perform this comparison in each band, given that both the data quality and empirical variability vary across wavebands (see Figure~\ref{fig:ibrm_lcs}). 

We find the RQ kernel to statistically perform the best in all bands, with the M-$\frac{1}{2}$ kernel performing only slightly poorer, and the SE kernel notably worse than the others. The NLML averaged across bands (lower indicates higher likelihood of the data given the model) is 27.8 for RQ, 29.8 for M-$\frac{1}{2}$, and 89.1 for SE. The MSE statistics more noticeably distinguishes the RQ and M-$\frac{1}{2}$ kernels, with the average MSE across bands being 1.2 for RQ, 1.5 for M-$\frac{1}{2}$, and 2.7 for SE. These results are in general agreement with previous GP modeling of AGN light curves \citep{Wilkins_2019, Griffiths_2021, Lewin_2022, Lewin_2023}. As an additional check, we compared the lags and their uncertainties that result from all four combinations of using RQ and M-$\frac{1}{2}$ to model each band and the \textit{UVW2} reference band. The lags are consistent within $1\sigma$ across kernel forms. At all frequencies, the lags agree within 5\% of each other on average, with the lag uncertainties agreeing within 10\%. 

The kernel comparison above was performed using the full 420-day light curves, in preparation for computing the lags presented in Section~\ref{subsubsec:campaign_lags}. Modeling the full campaigns with a single GP (per waveband) assumes consistent variability on both long and short timescales. In other words, a single set of hyperparameters is used to describe the variability of the entire campaign. In Section~\ref{subsubsec:epoch_lags}, however, we find that the UVOIR lags do vary significantly over the course of the campaign---splitting the light curves into three 140-day epochs produces lags in the second epoch over a factor of 2 shorter on average than those measured in the first and third epochs. This suggests non-stationarity in the source variability. We check how the modeling affects the epoch-resolved lags by re-computing the lags using a separate GP per epoch to better account for changes in the variability. The resulting lags are very consistent with those from using a single GP per waveband to model the entire campaign, as the lags agree within 5\% on average at all frequencies. We find similar agreement when using either the RQ kernel or the M-$\frac{1}{2}$ kernel to model the individual epochs.

\section{Recovering simulated time lags} \label{sec:sims}
\begin{figure}[t!]
    \centering
    \includegraphics[width=0.55\columnwidth]{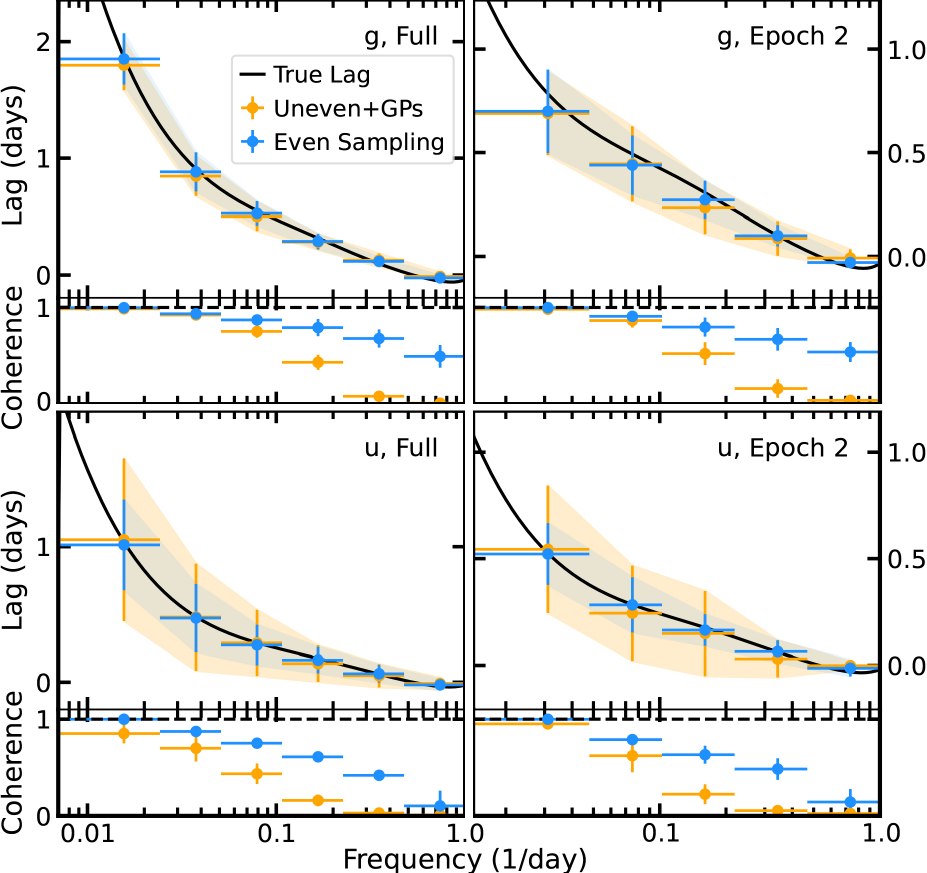}
    \caption{The lags and associated coherences computed with respect to the \textit{UVW2} band using simulated light curves in the \textit{g, u} bands (\textit{upper, lower}) for the full campaign (\textit{left panels}) and in Epoch~2 (\textit{right panels}). The injected/true lags indicated by the black curves are the final models used to reproduce the actual, measured lags in these bands. We use unevenly sampled light curves with time bins matching the observations, thus requiring GPs to compute the lags (orange). We compare these results to evenly sampled light curves (blue), from which the lags can be computed without GPs. Shaded regions indicate $1\sigma$ lag uncertainties.}
    \label{fig:sim_lfs}
\end{figure}

Here, we investigate how GP regression affects time lag recovery for our specific observations by simulating lags representative of our real measurements. In Section~\ref{sec:results}, we present the lags measured from 1) the full 420-day Swift and ground-based light curves and 2) three 140-day epochs. In both cases, we simulated the light curves of each waveband with lengths, means, and standard deviations matching those of the observations. The light curves were simulated using the method of \cite{Timmer_1995}, for which we assume a power-law PSD with a slope matching the best-fit slopes measured in Section~\ref{sec:results}. At frequencies lower than those probed in the lag analysis---0.007 day$^{-1}$ for the full-campaign and 0.014 day$^{-1}$ for the epoch-resolved lags--- the PSD slope is set to 0. In order to simulate time lags representative of our measurements, we convolve the light curves with the final (best-fit) impulse-response functions used to model the lags in Section~\ref{sec:modeling}; that is, the total model composed of the BLR and disk components. Just like the measured lags, a different model is used for each band and for each time range considered, resulting in 50 models used in total (12 UVOIR bands in 4 time ranges, and 2 X-ray bands for Epoch~2, the only epoch in which an X-ray/UV lag is detected). 

To most accurately reproduce the empirical data quality, the simulated light curves were then ``thinned," meaning that we considered only points in time that match the actual light curve in each band. In other words, the vector of time values should be roughly indistinguishable between the real and simulated light curves. In order to simulate white noise, each flux value was then re-drawn from a Gaussian distribution with a mean and standard deviation set by the flux and uncertainty, respectively, at each point.

After we generated these simulated light curves, we performed the same analysis on them as on the data presented in the paper. The variability in each band was then modeled by a GP, using the same kernel form (RQ) applied to the observations. The lags and their associated coherence values were then computed in the same frequency bins as the actual analysis. We compare these GP-recovered lags and coherences to those computed when the simulated light curves were instead binned to the average empirical sampling rate. In this case, Fourier techniques can be applied to the light curves immediately without GPs to compute the lags. As such, we use this comparison to gauge the effects of GPs on the lags, coherences, and their uncertainties, given the specific data quality. 

The simulated lag recovery for both sampling rates is presented in Figure~\ref{fig:sim_lfs}. The results agree with those found by \cite{Wilkins_2019, Lewin_2023}: we find, as expected, GP regression to more significantly affect the lag uncertainties and coherences than the sizes of the lags themselves. In all cases (regarding bands, frequencies, and treatment of the light curves), the GP-recovered lags are consistent within $1\sigma$ with those computed in the equal-sampling case. The true lag also lies within $1\sigma$ in all cases. For the lags computed using the full light curves, the GP-recovered lags are within 3\% on average from those computed from the light curves with equal sampling, although the uncertainties are larger by approximately 30\% on average. For the best- and worst-performing bands (\textit{g} and \textit{u}, respectively), the GP lags agree within 1\% and 5\% with those in the equal-sampling case. 

The coherences predicted by the simulated light curves after the use of GPs are generally consistent with those measured by the actual data. The effects of GPs on the coherence are minor at low frequencies ($<0.02~\text{day}^{-1}$), where the coherence is typically lower than that in the equal-sampling case by only $\sim$0.1. This is likely because the GP most easily preserves the overall (long-term) light curve shape and thus the correlated variability operating on long timescales. The coherence at high frequencies, however, is lower by 0.25--0.4 on average, depending on the band. At these high frequencies, the amplitude of the variability approaches that of the uncertainty on the data itself; thus, short timescale variability is produced in the gaps that is uncorrelated between light curve realizations of different wavebands.

In Section~\ref{subsubsec:campaign_lags}, we split the data into three epochs to probe changes in the lag over the course of the campaign. It was thus of particular importance to verify that applying GPs to different data qualities (although the data quality in Epochs~1 and 2 is similar in general, see Figure~\ref{fig:ibrm_lcs}) does not result in systematic changes in lag recovery. This does not appear to be the case, given that the GP-recovered lags in all bands and epochs agree with those that would have been measured had the data been equally sampled. Regardless, the true lag lies within $1\sigma$ for both sets of lags in all cases. Specifically, the GP-recovered lags lie within 2\% (Epochs~1 and 2) and 4\% (Epoch~3) on average from those computed in the equal-sampling case. Even in the poorest-data-quality scenario (the \textit{z} band in Epoch~3, as the \textit{u, $B_g$} bands are not used for Epoch~3 in the epoch-resolved analysis), the lags agree within 7\% at all frequencies. The uncertainties are larger on average by roughly 50\% in Epochs~1 and 2, and more significantly by 70\% in Epoch~3. 

\begin{deluxetable*}{lrrrrrr} \label{tab: full_lags}
\tablecaption{Time lags measured with respect to \textit{UVW2} (1928~\AA) using the full 420-day campaign}
\tablehead{\colhead{Filter} & \multicolumn{6}{c}{Lag (days) in Frequency Range} \\
\colhead{} & \colhead{0.01-0.02 day$^{-1}$} & \colhead{0.02-0.05 day$^{-1}$} & \colhead{0.05-0.11 day$^{-1}$} & \colhead{0.11-0.22 day$^{-1}$} & \colhead{0.22-0.48 day$^{-1}$} & \colhead{0.48-1.00 day$^{-1}$}}
\startdata
\textit{UVM2} & $0.37\pm0.32$ & $0.34\pm0.27$ & $-0.03\pm0.25$ & $-0.31\pm0.39$ & $-0.02\pm0.14$ & $-0.01\pm0.18$ \\
\textit{UVW1} & $0.67\pm0.32$ & $0.59\pm0.26$ & $0.07\pm0.24$  & $-0.09\pm0.53$ & $-0.01\pm0.11$ & $0.00\pm0.14$  \\
\textit{U}    & $1.55\pm0.35$ & $0.83\pm0.31$ & $0.19\pm0.36$  & $-0.11\pm0.65$ & $0.02\pm0.15$  & $0.00\pm0.19$  \\
\textit{u}    & $2.61\pm0.48$ & $1.12\pm0.72$ & $-0.09\pm0.53$ & $-0.24\pm1.29$ & $-0.04\pm0.67$ & $0.01\pm0.36$  \\
$B_g$ & $1.33\pm0.56$ & $0.66\pm0.46$ & $0.47\pm0.38$  & $-0.12\pm0.87$ & $-0.02\pm0.26$ & $0.00\pm0.25$  \\
$B$    & $1.51\pm0.45$ & $0.40\pm0.37$ & $0.44\pm0.32$  & $-0.07\pm0.37$ & $0.02\pm0.10$  & $0.00\pm0.13$  \\
$g$    & $2.47\pm0.28$ & $1.04\pm0.24$ & $0.52\pm0.19$  & $0.06\pm0.50$  & $-0.01\pm0.09$ & $0.00\pm0.08$  \\
$V_g$ & $3.06\pm0.33$ & $1.45\pm0.27$ & $0.59\pm0.29$  & $0.01\pm0.56$  & $-0.03\pm0.10$ & $-0.01\pm0.11$ \\
$V$    & $2.68\pm0.58$ & $1.99\pm0.38$ & $0.57\pm0.47$  & $0.06\pm0.48$  & $0.03\pm0.19$  & $0.02\pm0.23$  \\
$r$    & $4.13\pm0.26$ & $2.15\pm0.22$ & $0.89\pm0.24$  & $0.18\pm0.47$  & $0.02\pm0.10$  & $0.00\pm0.11$  \\
$i$    & $4.67\pm0.31$ & $3.36\pm0.25$ & $1.32\pm0.30$  & $0.26\pm0.52$  & $0.04\pm0.11$  & $0.01\pm0.14$  \\
$z$    & $5.50\pm1.33$ & $3.54\pm1.40$ & $1.93\pm3.01$  & $0.05\pm1.77$  & $0.02\pm0.83$  & $0.00\pm0.41$  \\
\enddata
\end{deluxetable*}

\begin{deluxetable*}{lrrr} \label{tab: epochresolved_lags}
\tablecaption{Time lags measured with respect to \textit{UVW2} (1928~\AA) at low frequencies ($0.014-0.048$ day$^{-1}$, equivalent to timescales of $20-70$ days) when evenly dividing the campaign into three epochs.}
\tablehead{\colhead{Filter} & \multicolumn{3}{c}{Lag (days) in Time Range} \\
\colhead{} & \colhead{Epoch 1} & \colhead{Epoch 2} & \colhead{Epoch 3}}
\startdata
\textit{UVM2} & $0.32\pm0.26$ & $0.40\pm0.29$ & $0.46\pm0.33$\\
\textit{UVW1} & $0.48\pm0.25$ & $0.60\pm0.27$ & $0.17\pm0.31$\\
\textit{U}    & $1.26\pm0.27$ & $0.80\pm0.29$ & $1.19\pm0.40$\\
\textit{u}    & $1.69\pm0.34$ & $0.80\pm0.53$ & $1.48\pm1.24$\\
$B_g$  & $1.66\pm0.35$ & $0.94\pm0.29$ & $1.59\pm1.57$\\
$B$    & $1.22\pm0.29$ & $0.29\pm0.33$ & $1.04\pm0.40$\\
$g$    & $1.90\pm0.17$ & $0.88\pm0.20$ & $0.82\pm0.43$\\
$V_g$  & $2.24\pm0.22$ & $0.84\pm0.26$ & $1.78\pm0.44$\\
$V$    & $2.05\pm0.37$ & $0.92\pm0.31$ & $2.01\pm0.54$\\
$r$    & $2.59\pm0.19$ & $1.32\pm0.20$ & $2.76\pm0.35$\\
$i$    & $3.13\pm0.20$ & $1.87\pm0.20$ & $3.38\pm0.45$\\
$z$    & $4.29\pm0.41$ & $2.19\pm0.28$ & $6.09\pm3.43$\\
\enddata
\end{deluxetable*}

\begin{figure*}[t!]
    \centering
    \includegraphics[width=\textwidth]{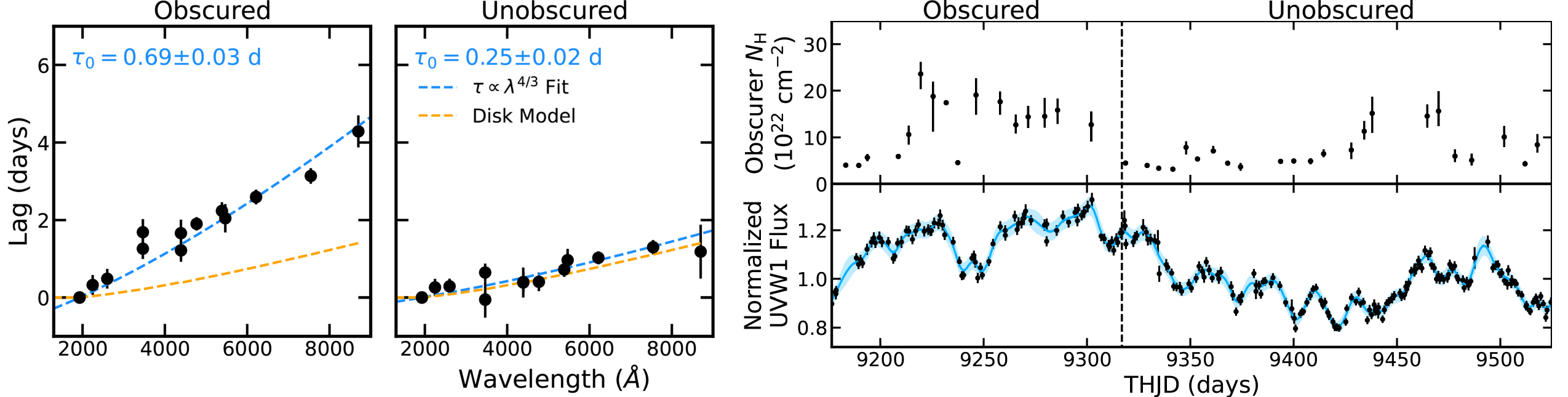}
    \caption{\textit{Left}: Lags as a function of wavelength at frequencies of $0.014-0.048~\text{day}^{-1}$, equivalent to timescales of $20-70~\text{days}$ computed during periods of relatively high and low obscurer column density. The lags are fit with a $\tau \propto \lambda^{4/3}$ relation (dashed blue line), with best-fit normalization values ($\tau_0$) shown. The expected lag-wavelength relation of the disk reprocessing model is shown in orange. The disk model lies within $1\sigma$ from the $\tau \propto \lambda^{4/3}$ best fit for the unobscured lags. \textit{Upper right:} Obscurer column density over the course of the campaign from the NICER spectral analysis in Paper~III \citep{Partington_2023}. \textit{Lower right}: UVW1 light curve from Figure~\ref{fig:ibrm_lcs}. Dashed lines demarcate the two time ranges used to compute the lags. The obscured time range is the same as the original Epoch~1 used in the epoch-resolved analysis. The unobscured time range combines the original Epoch~2 with the (unobscured) first half of the original Epoch~3. We do not use the last 75 days of the campaign (the second half of the original Epoch~3) when the column density is high in this analysis, and thus it is not shown. The lags shown here can be found in Table~\ref{tab: epochresolved_lags_highlow}.}
    \label{fig:lws_longer}
\end{figure*}

\begin{deluxetable*}{lrr} \label{tab: epochresolved_lags_highlow}
\tablecaption{Time lags measured with respect to \textit{UVW2} (1928~\AA) at low frequencies ($0.014-0.048$ day$^{-1}$, equivalent to timescales of $20-70$ days) computed during periods of relatively high and low obscurer column density, as shown in Figure~\ref{fig:lws_longer}.}
\tablehead{\colhead{Filter} & \multicolumn{2}{c}{Lag (days) in Time Range} \\
\colhead{} & \colhead{Obscured} & \colhead{Unobscured}}
\startdata
\textit{UVM2} & $0.32\pm0.26$ & $0.26\pm0.22$\\
\textit{UVW1} & $0.48\pm0.25$ & $0.29\pm0.20$\\
\textit{U}    & $1.26\pm0.27$ & $0.64\pm0.23$\\
\textit{u}    & $1.69\pm0.34$ & $-0.05\pm0.46$\\
$B_g$  & $1.66\pm0.35$ & $0.40\pm0.26$\\
$B$    & $1.22\pm0.29$ & $0.38\pm0.39$\\
$g$    & $1.90\pm0.17$ & $0.40\pm0.24$\\
$V_g$  & $2.24\pm0.22$ & $0.72\pm0.19$\\
$V$    & $2.05\pm0.37$ & $0.96\pm0.29$\\
$r$    & $2.59\pm0.19$ & $1.02\pm0.17$\\
$i$    & $3.13\pm0.20$ & $1.30\pm0.18$\\
$z$    & $4.29\pm0.41$ & $1.18\pm0.70$\\
\enddata
\end{deluxetable*}

\bibliography{reference}{}
\bibliographystyle{aasjournal}

\end{document}